\documentclass[]{spie}  %

\usepackage{amsmath,amsfonts,amssymb}
\usepackage{graphicx}
\usepackage{multicol}
\usepackage{multirow}
\usepackage{booktabs}
\usepackage[colorlinks=true, allcolors=blue]{hyperref}
\usepackage{subcaption}
\usepackage{paralist}
\usepackage{lineno}
\usepackage{xcolor}

\newcommand{\norm}[1]{\left\lVert#1\right\rVert}
\newcommand{\Dgan}{\boldsymbol{D}}
\newcommand{\Ggan}{\boldsymbol{G}}
\newcommand{\mb}{\boldsymbol{m}}
\newcommand{\pb}{\boldsymbol{p}}
\newcommand{\dint}{\mathrm{d}}

\title{X-ray Photon-Counting Data Correction through Deep Learning}

\author[a]{Mengzhou Li}
\author[b]{David S. Rundle}
\author[a]{Ge Wang$^*$}
\affil[a]{Biomedical Imaging Center, Department of Biomedical Engineering, Center for Biotechnology \& Interdisciplinary Studies, Rensselaer Polytechnic Institute, Troy, NY USA }
\affil[b]{JairiNovus Technologies Ltd., 119 Heartland Drive, Butler, PA, USA}

\authorinfo{$^*$ Corresponding author: Ge Wang. E-mail: wangg6@rpi.edu.}

\begin{document} 
\maketitle

\begin{abstract}
X-ray photon-counting detectors (PCDs) are drawing an increasing attention in recent years due to their low noise and energy discrimination capabilities. The energy/spectral dimension associated with PCDs potentially brings great benefits such as for material decomposition, beam hardening and metal artifact reduction, as well as low-dose CT imaging. However, X-ray PCDs are currently limited by several technical issues, particularly charge splitting (including charge sharing and K-shell fluorescence re-absorption or escaping) and pulse pile-up effects which distort the energy spectrum and compromise the data quality. Correction of raw PCD measurements with hardware improvement and analytic modeling is rather expensive and complicated. Hence, here we proposed a deep neural network based PCD data correction approach which directly maps imperfect data to the ideal data in the supervised learning mode. In this work, we first establish a complete simulation model incorporating the charge splitting and pulse pile-up effects. The simulated PCD data and the ground truth counterparts are then fed to a specially designed deep adversarial network for PCD data correction. Next, the trained network is used to correct separately generated PCD data. The test results demonstrate that the trained network successfully recovers the ideal spectrum from the distorted measurement within $\pm6\%$ relative error. Significant data and image fidelity improvements are clearly observed in both projection and reconstruction domains.
    
\end{abstract}


\section{INTRODUCTION}
\label{sec:intro}  %

Photon-counting detectors (PCDs) have been popular in astronomy\cite{barbieri2009aqueye,verhoeve2008photon}, communication\cite{robinson2006781,buck2019photon}, material science\cite{bergamaschi2009photon}, medical imaging\cite{kalender2017technical, symons2016low}, optical imaging\cite{becker2004fluorescence,isbaner2016dead}, and other areas.
In the medical imaging field, X-ray PCDs are drawing {an }increasing attention from both industrial and academic sides in recent years\cite{leng2018150, taguchi2018spatio, amma2019assessment,hansson2019high,sossin2018experimental}. Besides the low noise feature, X-ray PCDs count photons into different energy bins in sharp contrast to energy-integrating detectors (EIDs). Thus, PCDs introduce {a unique} spectral dimension, {data in multiple} energy bins, relative to {signals in a single energy bin from} conventional EIDs. This energy discrimination ability can significantly facilitate material decomposition tasks, and solve or alleviate beam hardening and metal artifact problems. In addition, compared to EIDs which have small weights on low-energy photons, PCDs weigh optimally low-energy and high-energy photons which increases the contrast and {reduces} radiation dose. The advantages of PCDs over EIDs have been demonstrated in many quantitative CT tasks, and the use of PCDs will also pave the way for novel applications like simultaneous multi-agent imaging and molecular CT. 

However, the current {performance of PCDs} is far from being flawless, and several technical issues still hinder wider applications of PCDs; e.g., counting speed limit, spatial non-uniformity, and spectrum distortion. Especially, in clinical CT which utilizes a high flux of X-rays, the PCD {data quality} will be greatly degraded by {the pulse pileup effect} resulting in a count loss and a spectrum distortion. In more severe cases, the PCD gets polarized due to the sensor impurities and does not even count photons. Smaller pixels could be used to address the counting rate problem, and thus to raise the {maximum} counting rate of the detector effectively {and at the same time }refine the {CT image} spatial resolution. Nonetheless, small pixels suffer from the charge splitting effect which includes the charge sharing between neighboring pixels and the K-shell fluorescence escape from the pixel, which might cause a serious spectrum distortion toward the low energy direction. Vendors designed various anti-charge sharing application-specific integrated circuits (ASICs), compensating for the spectrum distortion caused by charge sharing at the cost of complex circuitry. The complex ASICs, on the other hand, put a limit on the counting speed. Consequently, pulse pileup and charge splitting are the two main challenges that PCDs are facing, which contribute most to the spectrum distortion. Another issue is the spectral and intensity-dependent spatial non-uniformity caused by the charge trapping and charge steering effects as well as the variation in electronic properties between pixelated ASICs, which is less of a problem and can be overcome through careful calibrations \cite{wang2008uniformity, getzin2019non}.

Despite the efforts by PCD manufactures to correct pulse pileup and charge sharing problems with hardware solutions aspects\cite{fu2018apparatus,tkaczyk2010high}, many researchers investigated less expensive correction algorithms \cite{ponchut2008correction, ding2012image, taguchi2013vision}. The general idea behind many of the correction algorithms is to build the forward model describing the charge splitting and pulse pileup induced spectral distortions, and then use an iterative method to estimate the real spectrum by maximizing the likelihood for the measured data and modeled output. The key to the success of this kind of algorithm is the accuracy of the forward model. Monte Carlo simulation could provide accurate results but at a slow speed, and many scholars devised fast models to depict the PCD count statistics\cite{wang2011pulse, tanguay2015detective, xu2014cascaded}. Taguchi extensively studied various spectrum distortion mechanisms including charge splitting and pulse pileup\cite{taguchi2013vision}, and built a cascaded PCD open access simulator with correlated Poisson noise recently but without pulse pileup\cite{taguchi2016spatio, taguchi2018spatio}. Instead of using approximate prediction of pulse pile-up distortion, Philips scientists proposed an exact analytical prediction model \cite{roessl2016fourier} based on the frequency calculation of level crossing of shot noise processes \cite{bierme2012fourier}, and presented its impressive agreement with measured data on 2019 CERN workshop. Even with that progress in PCD modeling, it is still hard to obtain accurate correction on real PCDs due to the complexity of modeling various physical effects inside the PCDs.

Recently, deep learning has been a hot topic in the imaging field and was applied successfully in many challenging tasks with impressive results, such as image denoising \cite{shan20183d, yang2017low, you2018structurally}, image super-resolution \cite{you2019ct, li2020deep}, compressive imaging \cite{yao2019net}, motion correction, and image reconstruction \cite{zhu2018image}. Deep learning is an end-to-end data-driven approach, which does not rely on an explicit physical model as usually required by traditional model-based methods, and instead it can automatically learn hidden rules through a complex network during the training process with paired data. Taking advantage of this supervised learning mode, we can circumvent the complex model of the PCD response and directly map the distorted raw data to the ground truth spectrum. Along the direction, Badea corrected the spectral distortion with a two-hidden-layer fully connected neural network \cite{touch2016neural}, and made improvements in material concentration quantification compared to that with the uncorrected measurements. But due to the limited representation ability of the shallow network, the correction results were not ideal and there is large room for further improvement. 

In this paper, we propose a Deep Neural Network (DNN) for PCD data corretion. Our method directly and simultaneously corrects the spectral distortion and removes noise within a unified DNN framework. The general idea is as follows. First, we generate numerous phantoms with various shapes and compositions with 3D printing or liquid tissue surrogate\cite{wang2019deformable} techniques to generate real raw data. Then with the known geometry and material composition, we compute ideal data aided by the linear attenuation coefficient (LAC) database from the National Institute of Standards and Technology (NIST) and X-ray source spectrum simulation software. Finally, the DNN can be trained to map the distorted noisy projections to the non-distorted noise-free ground truth projections, and the trained network with the denoising and spectrum-rectifying abilities can be applied for correction of PCD projections acquired in similar settings. In principle, utilizing the proposed method we are able to calibrate the PCD-based reconstruction results to the standard of the NIST database, enabling quantitative CT. In this pilot study, instead of collecting real distorted data, we used synthesized distorted PCD data which includes charge splitting, pulse pileup and Poisson noise for feasibility demonstration. Although the PCD simulation model may not be absolutely perfect, the simulated data should be realistic enough for our purpose.

\section{Data and Models}\label{sec:Method}
The general goal is to correct the PCD detection data to the ideal data with spectral fidelity. First, the ideal spectrum of an attenuated X-ray beam is calculated from the synthesized material phantom and a close-to-real X-ray source spectrum generated with a mature software tool. Then, the workflow of the PCD detection model is described covering charge splitting (charge sharing and fluorescence re-absorption or escaping), Poisson noise, and pulse pileup. Finally, our deep neuron network is introduced for PCD distortion correction.

\subsection{Phantom and geometry settings}
\begin{figure}[htbp]
	\centering
	\includegraphics[width=0.5\textwidth]{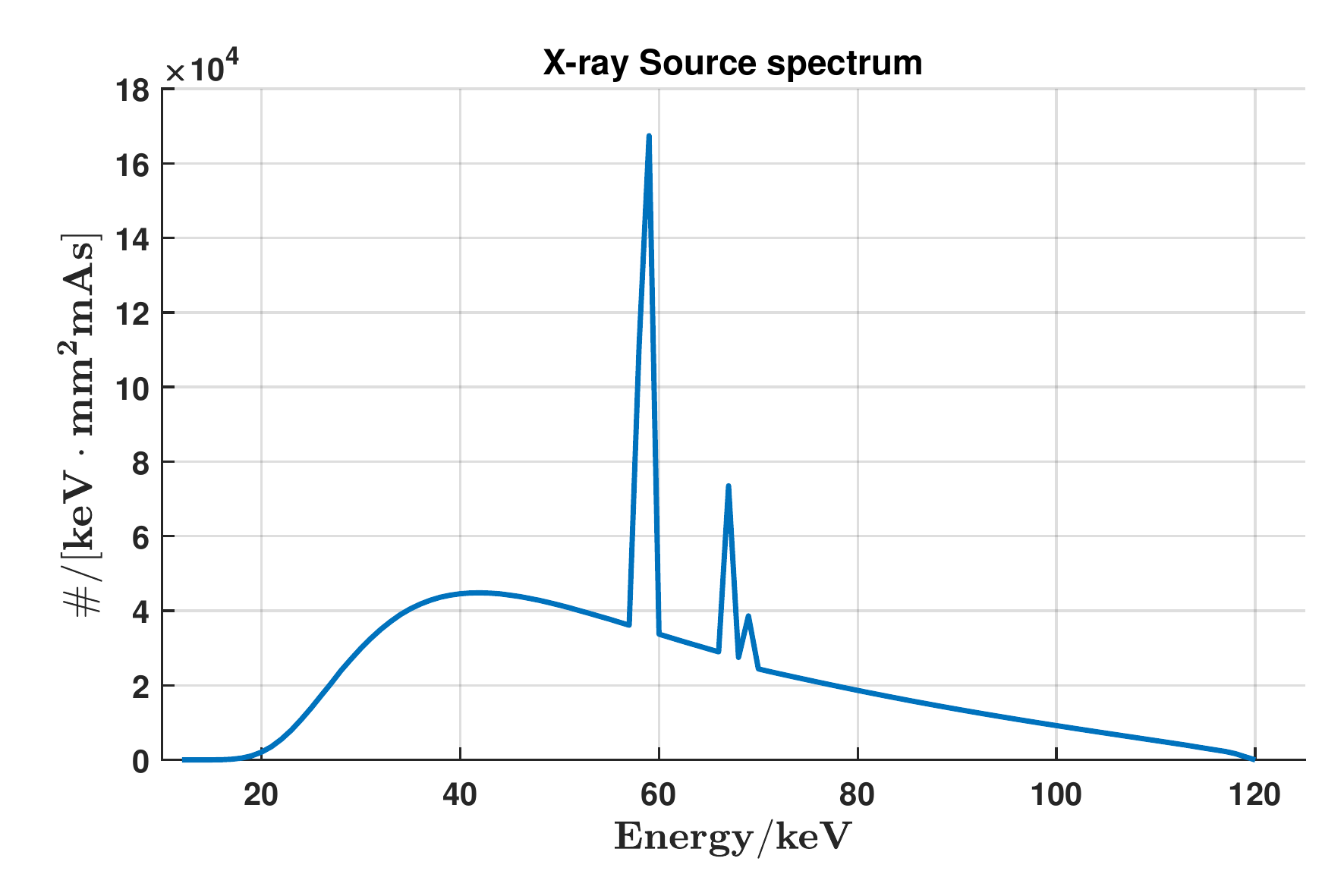}
	\caption{X-ray spectrum at $1 m$ from the source with filtration comprised of $0.8 mm$ Be, $1 mm$ Al and $0.11 mm$ Cu.}
	\centering
	\label{fig:xrayspectrum}
\end{figure}

To correct spectral distortion of PCDs using a data driven method, a large amount of data are required for network training. A polychromatic source and energy dependent attenuation curves are two ingredients for PCD data generation. The X-ray source spectrum was simulated with \textit{SpekCalc}\cite{poludniowski2009spekcalc} operated in a diagnostic energy range. The simulated spectrum ranges from $12 keV$ to $120 keV$ and has a resolution of $1 keV$. The distance between the phantom and the source was set to 1 meter, and the default filtration consisting of $0.8 mm$ Beryllium, $1 mm$ Aluminium and $0.11 mm$ Copper. The output spectrum is shown in Fig. \ref{fig:xrayspectrum}. 

\begin{figure}
   \begin{minipage}[b]{.49\linewidth}
     \centering
     \begin{minipage}[b]{.49\linewidth}
        \includegraphics[width=1\textwidth]{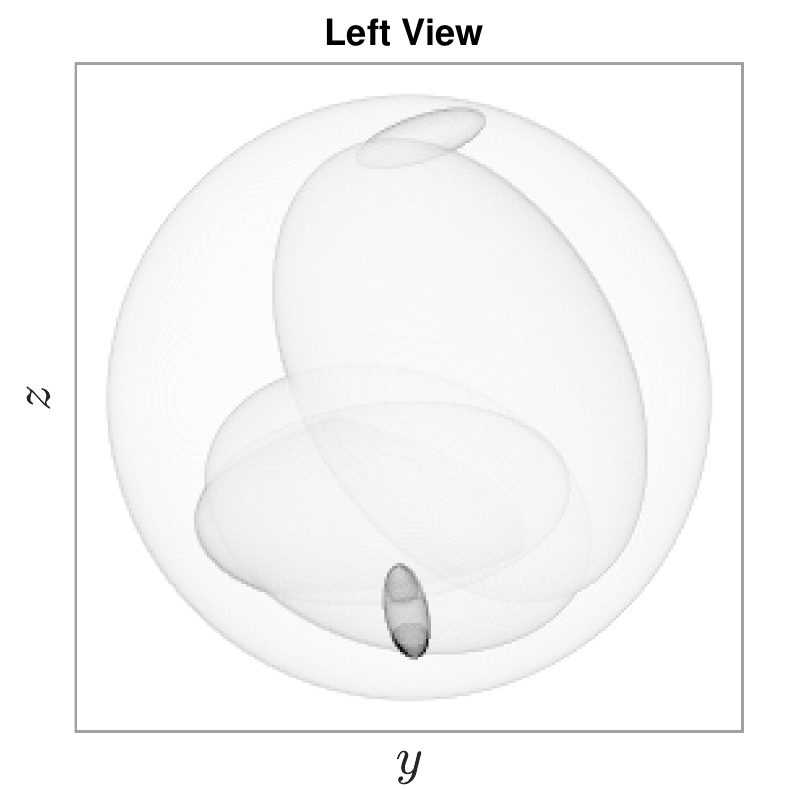}
        \includegraphics[width=1\textwidth]{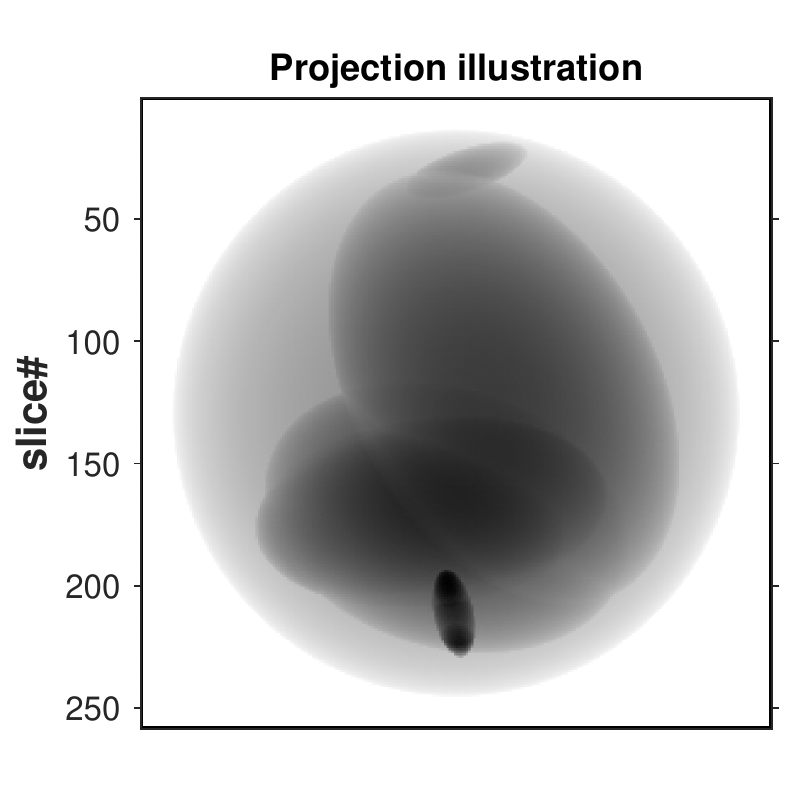}
     \end{minipage}
     \hfill
     \begin{minipage}[b]{.49\linewidth}
        \includegraphics[width=1\textwidth]{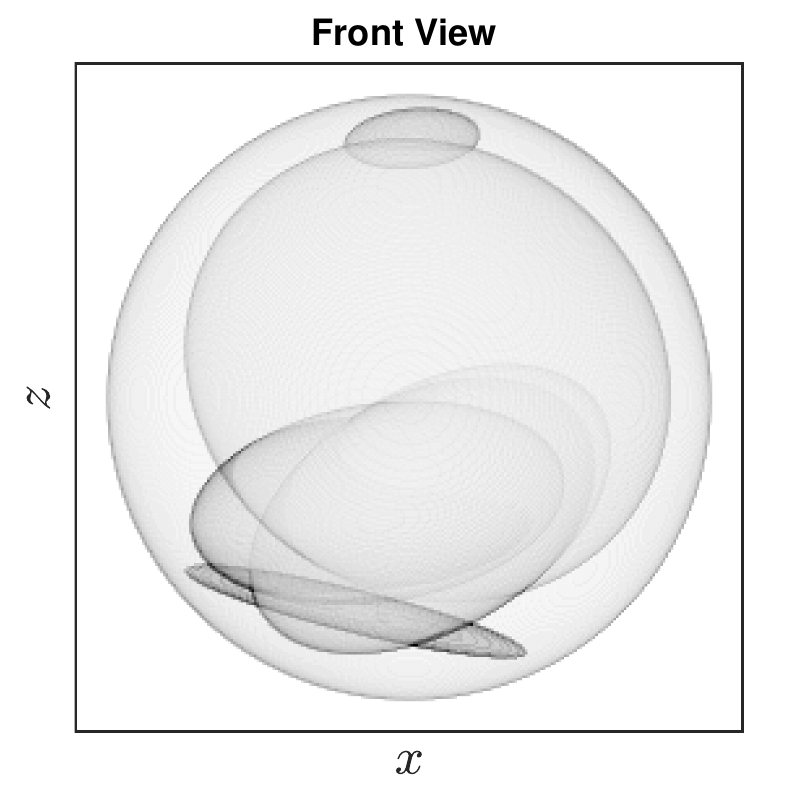}
        \includegraphics[width=1\textwidth]{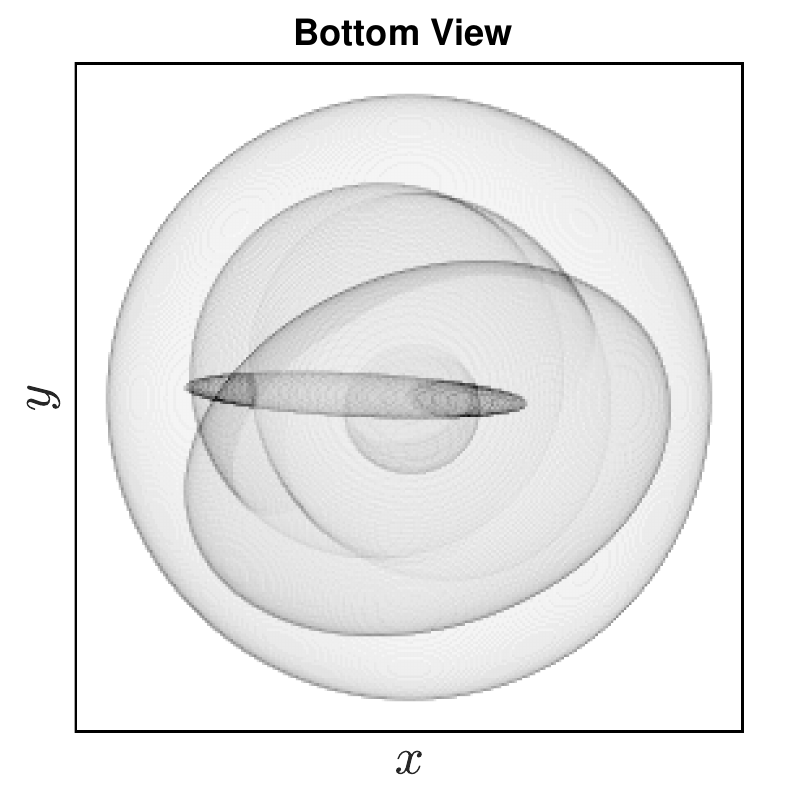}
     \end{minipage}
     \subcaption{Phantom geometry}\label{fig:PhantomG}
   \end{minipage}
   \hfill
   \begin{minipage}[b]{.45\linewidth}
     \centering
     \begin{minipage}[b]{.52\linewidth}
        \includegraphics[width=1\textwidth]{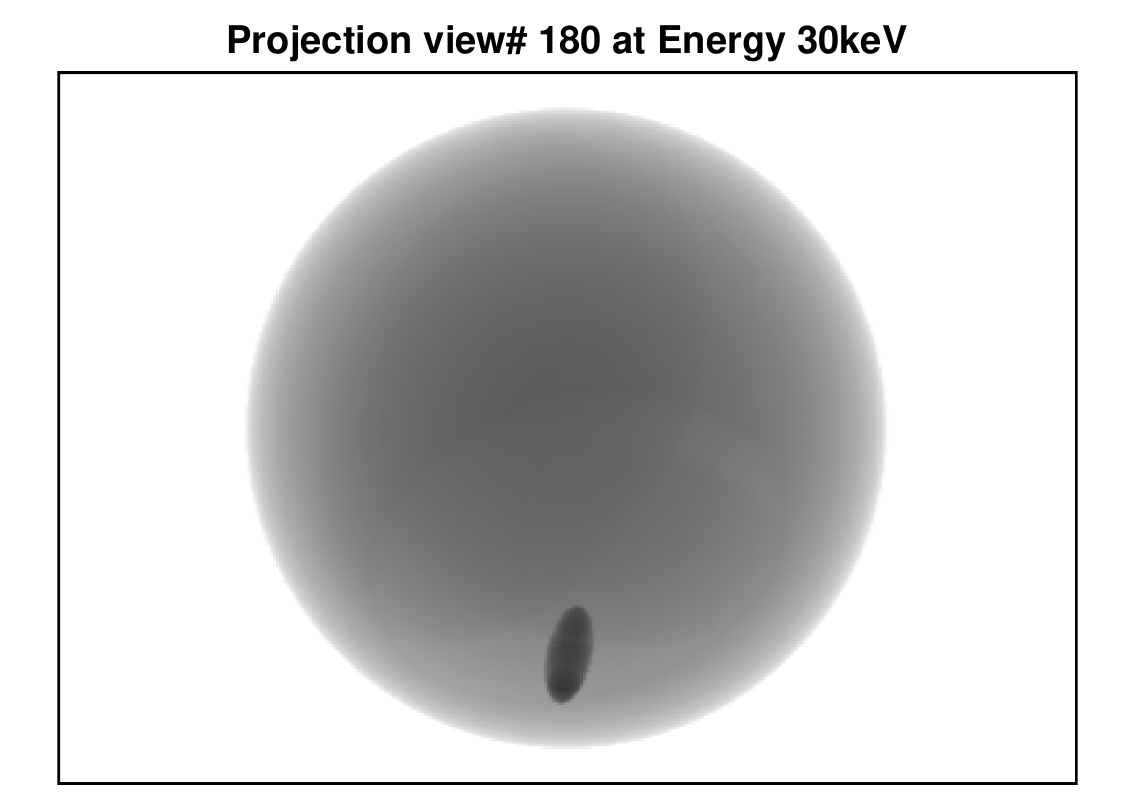}
        \includegraphics[width=1\textwidth]{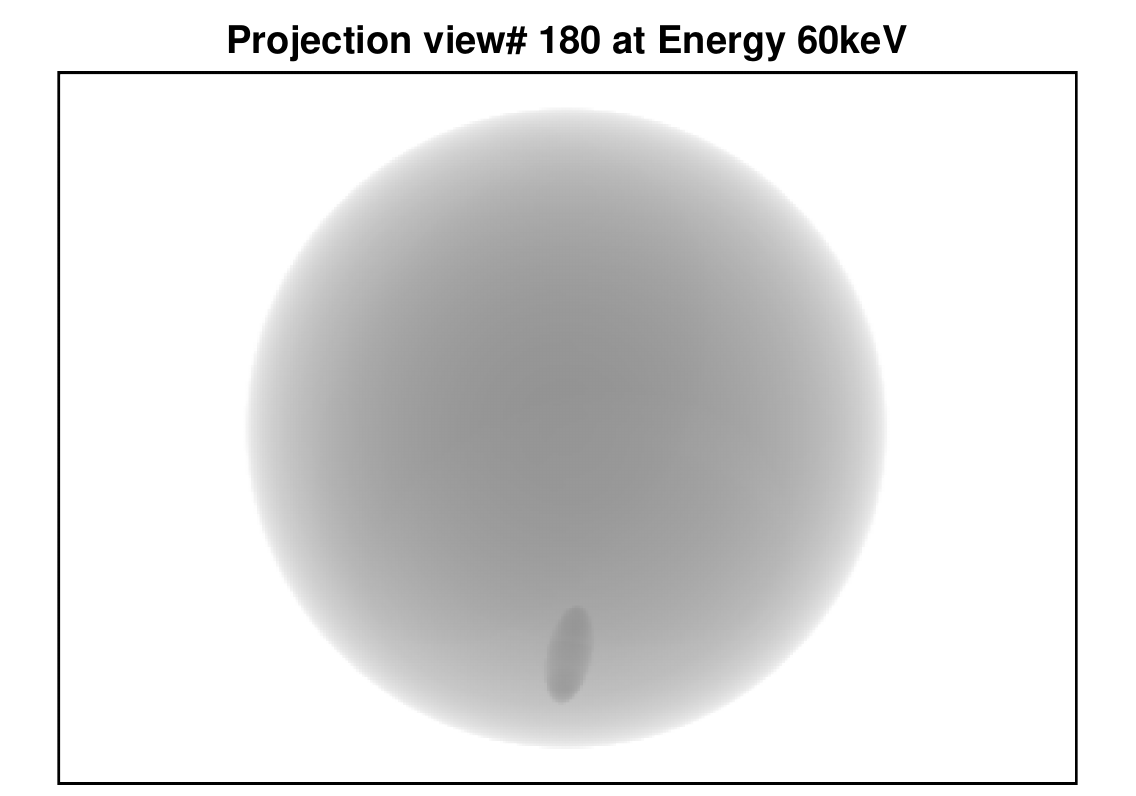}
        \includegraphics[width=1\textwidth]{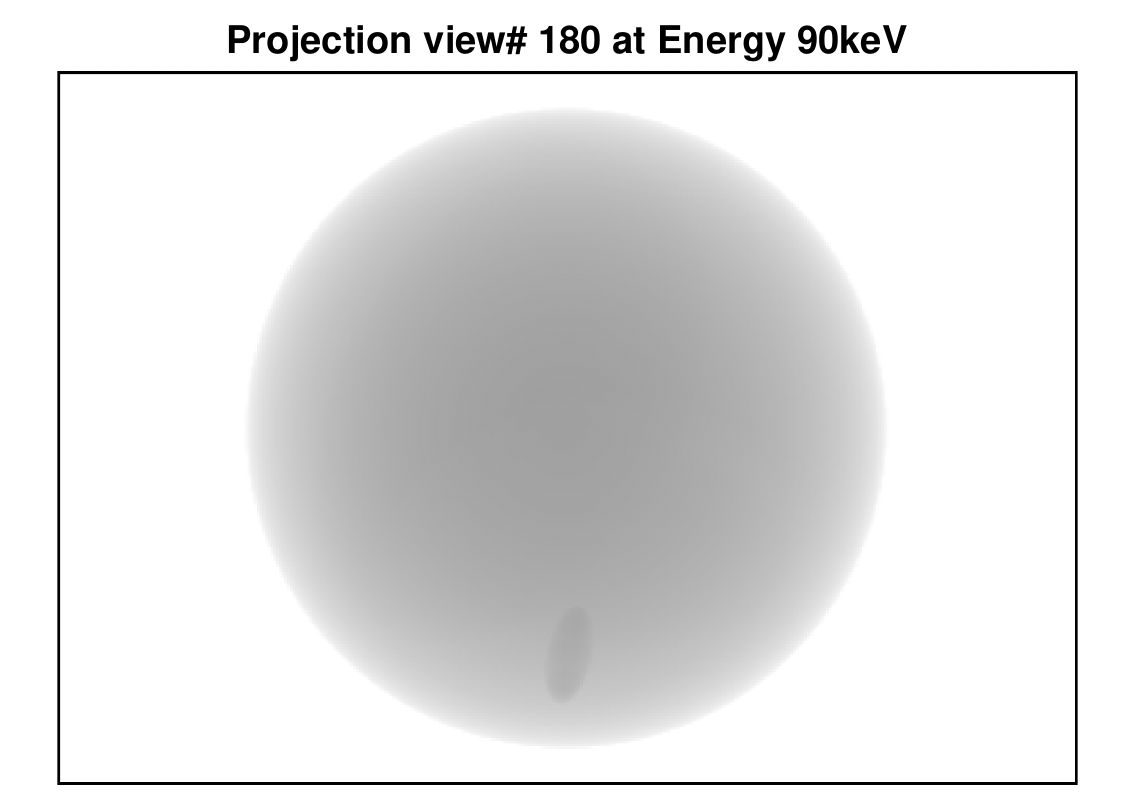}
     \end{minipage}
     \hfill
     \begin{minipage}[b]{.38\linewidth}
        \includegraphics[width=1\textwidth]{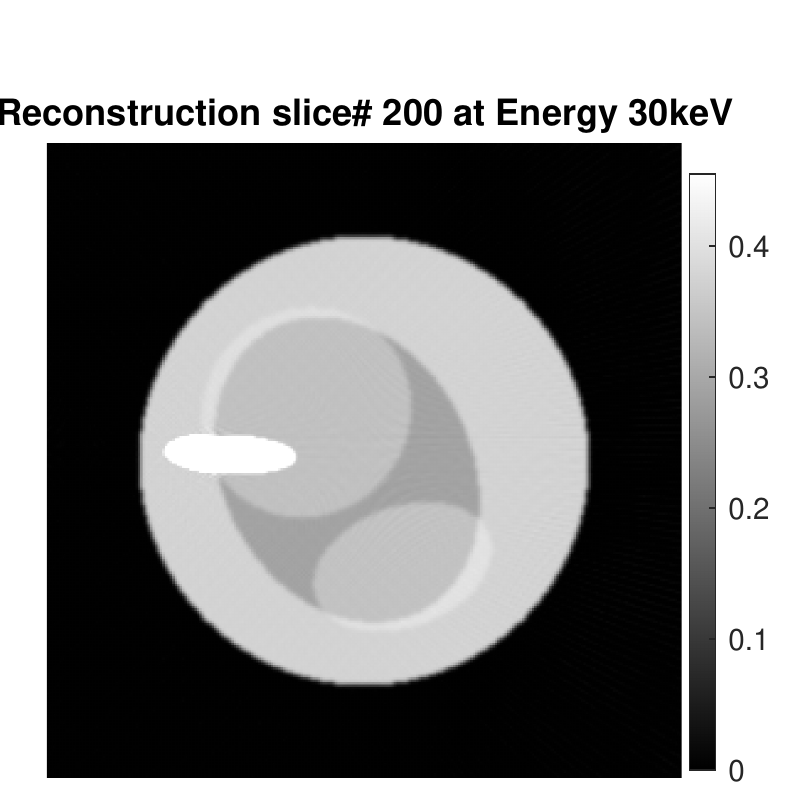}
        \includegraphics[width=1\textwidth]{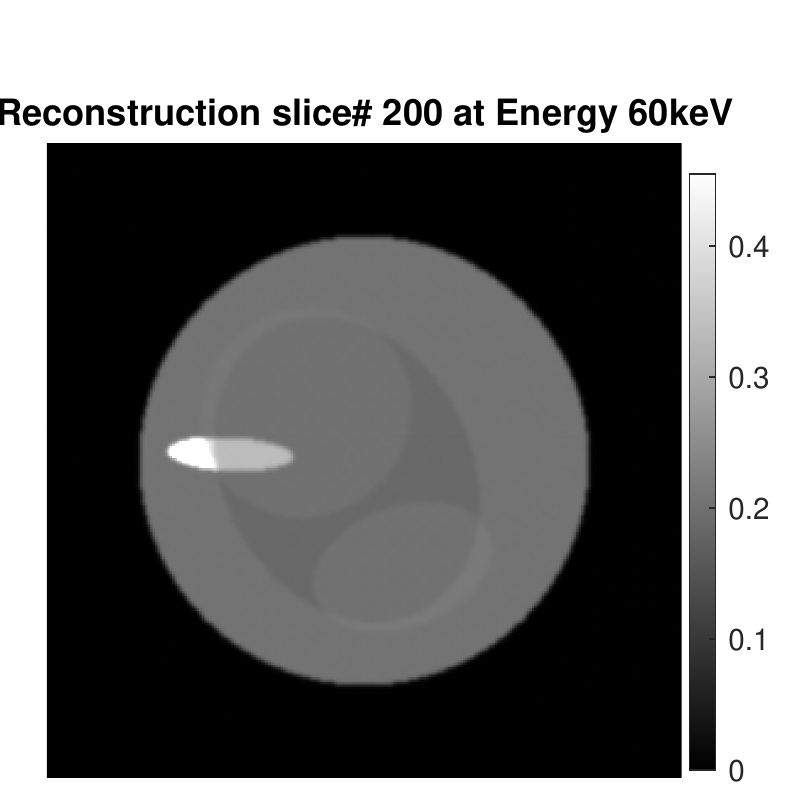}
        \includegraphics[width=1\textwidth]{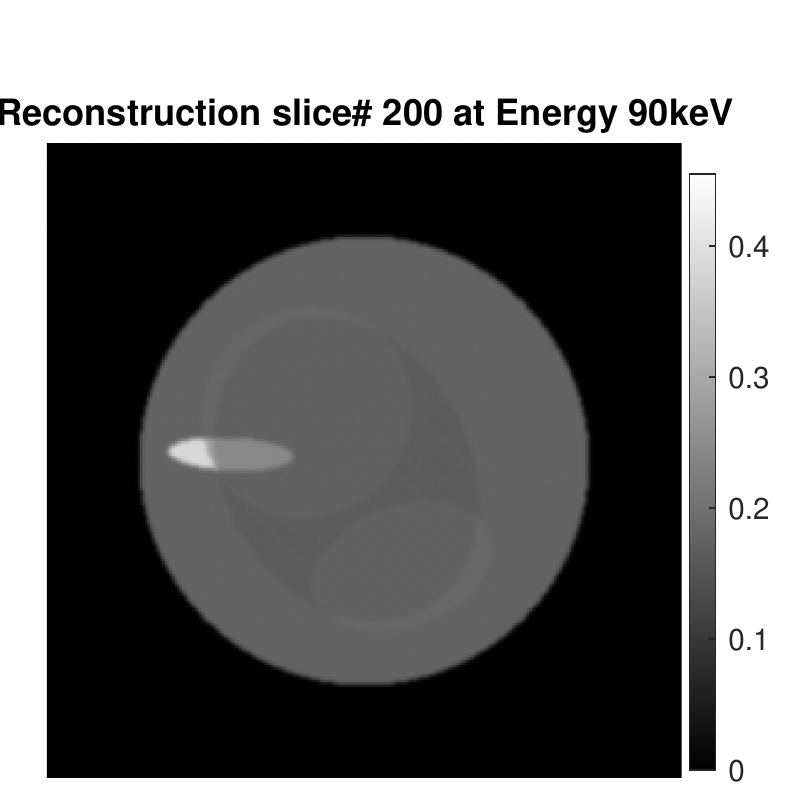}
     \end{minipage}
     \subcaption{Projections and Reconstruction}\label{fig:PhantomRecon}
   \end{minipage}
   \caption{Material phantom geometry and spectrally dependent attenuation. (a) Phantom volume visualization from different views; and (b) projections at the $180^\circ$ angle view at different photon energies and corresponding reconstructions of the $200th$ slice.}
   \label{fig:PhatomIllustration}
\end{figure}

As for representative objects to be scanned, a 3D world of Shepp-Logam phantoms with random shapes and multiple material compositions were used for data generation. Each phantom was a 256*256*256 cubic with voxel size of $0.11^3mm^3$, and 5 ellipsoids of different materials were located completely inside a sphere of radius $1.28 cm$ centered in the cube. The geometry of each ellipsoid was randomly specified by the following parameters: center position in spherical coordinates $(r,\theta,\theta_{z})$, semi-axes $(a, r, r)$, and an orientation angle around the $z$ axis $\phi_{z}$. Five material types (soft tissue, adipose tissue, brain grey and white matter, blood and cortical bone) were assigned to the five ellipsoids. The ellipsoids can overlap with each other, and the material composition of an overlapped pixel is assigned with the equal-volume mixture of the involved ellipsoids. The gaps between the ellipsoids and the sphere boundary was filled with water, and the space outside the sphere was treated as air. We also constrained the size and roundness of the ellipsoids according to their material types. The roundness (represented by the relative magnitude difference between $a$ and $r$) of soft tissue, adipose tissue, brain tissue and bone was gradually decreasing; i.e., bone was in a bar shape while soft tissue was close to a sphere. The roundness of blood was randomly selected. Similarly, the ellipse size (represented by the magnitude of $a$ and $r$) was also decreasing in the same order except for bone which was relatively large. The geometry of an example phantom is illustrated in Fig. \ref{fig:PhantomG}.

\begin{figure}
   \begin{minipage}[b]{.45\linewidth}
     \centering
     \includegraphics[width=1.0\textwidth]{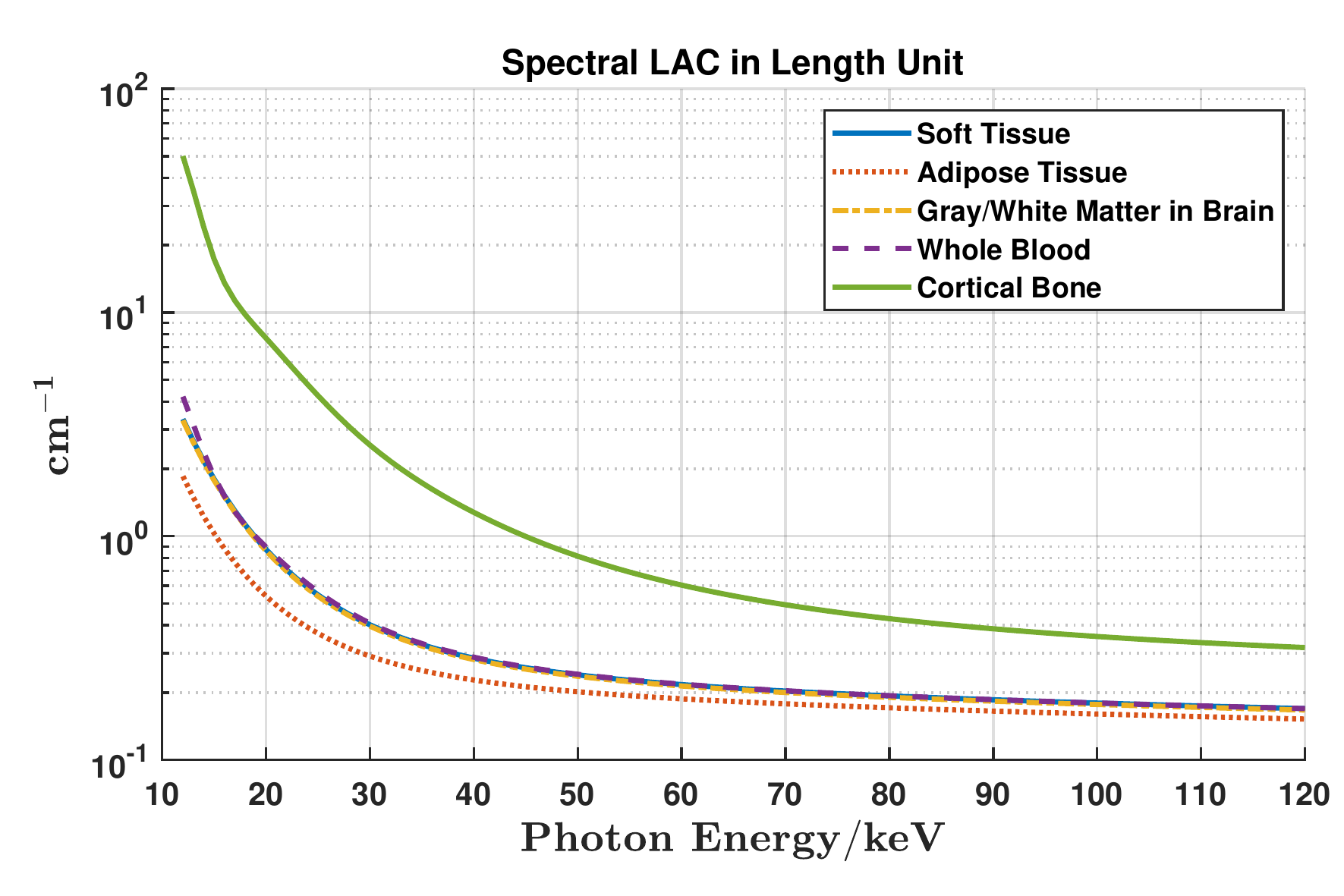}
     \subcaption{LAC in length unit}
   \end{minipage}
   \hfill
   \begin{minipage}[b]{.45\linewidth}
     \centering
     \includegraphics[width=1.0\textwidth]{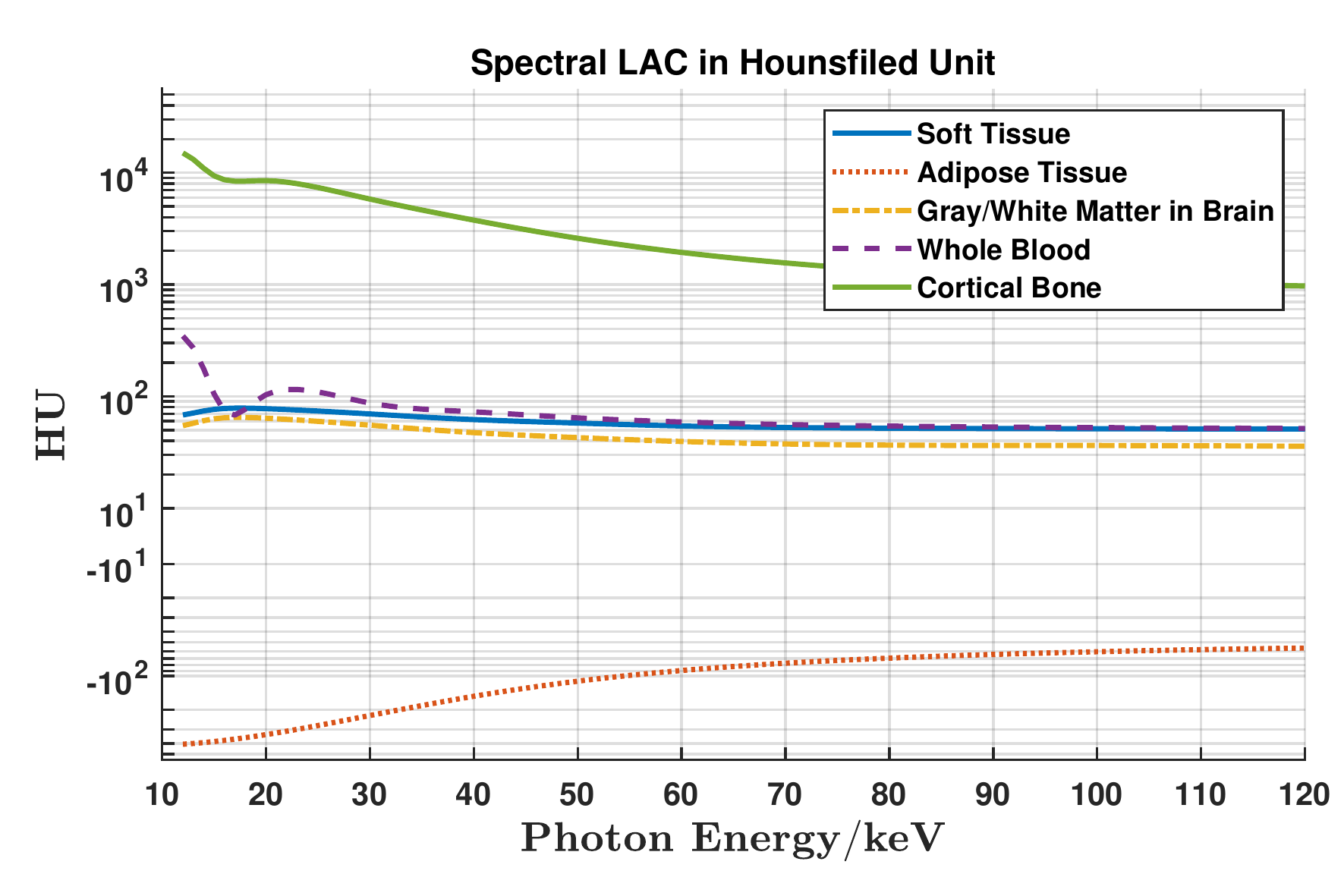}
     \subcaption{LAC in Hounsfield Unit}
   \end{minipage}
   \caption{Spectral LAC curves of different human materials. For better illustration, the y axis is on a logarithm scale. The LAC values in the Hounsfield unit were calculated with reference to the water LAC at the same X-ray photon energy.}
   \label{fig:LACs}
\end{figure}

The spectral LAC data for those representative human tissues and other materials are from the NIST database\cite{icru1989tissue}, as shown in Fig. \ref{fig:LACs}. The LAC values at energy points between the NIST data points were interpolated via log-log cubic-spline fitting, which is the same method used by NIST for interpolation from measured and calculated data points\cite{berger1987xcom}. 

\begin{figure}
   \begin{minipage}[b]{.45\linewidth}
     \centering
     \includegraphics[width=1.0\textwidth]{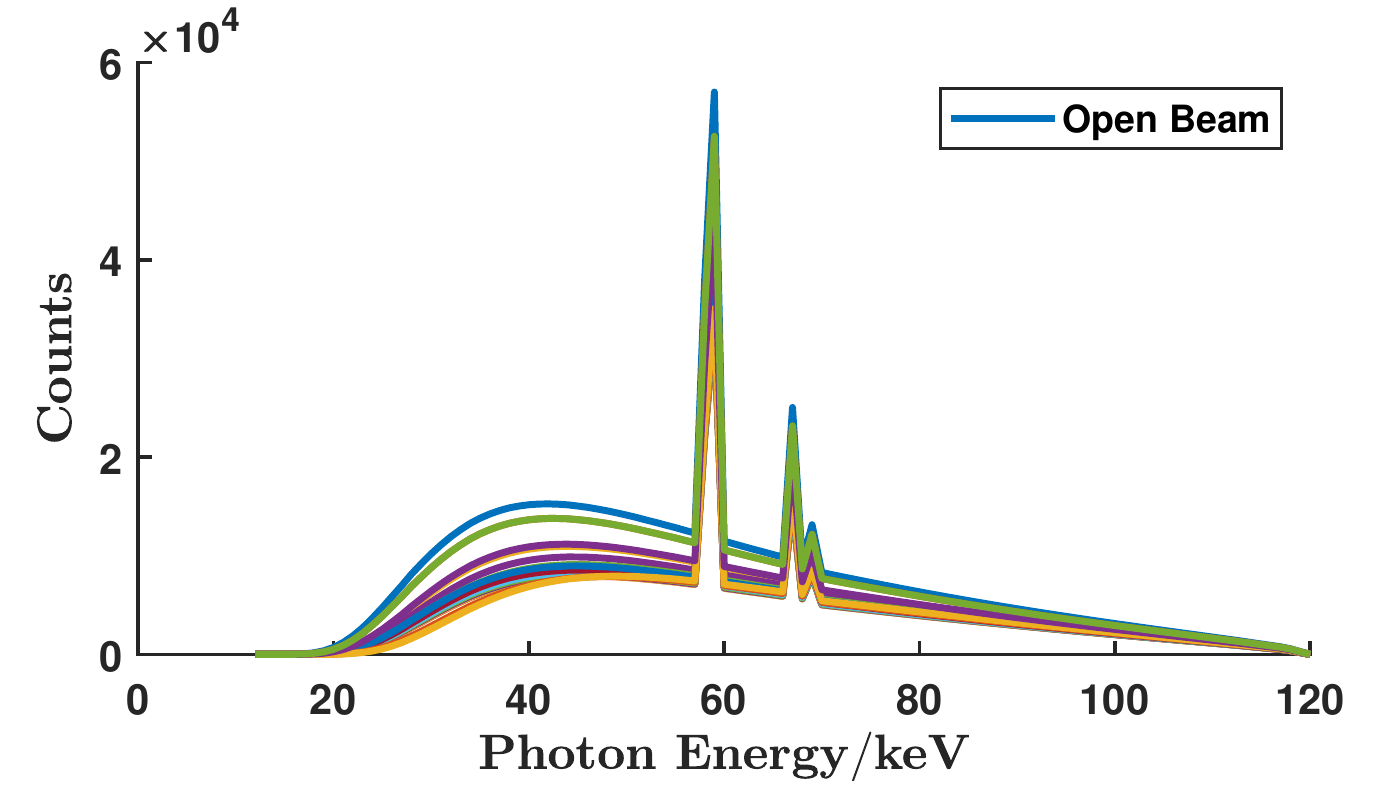}
     \subcaption{Spectra in counts}
   \end{minipage}
   \hfill
   \begin{minipage}[b]{.45\linewidth}
     \centering
     \includegraphics[width=1.0\textwidth]{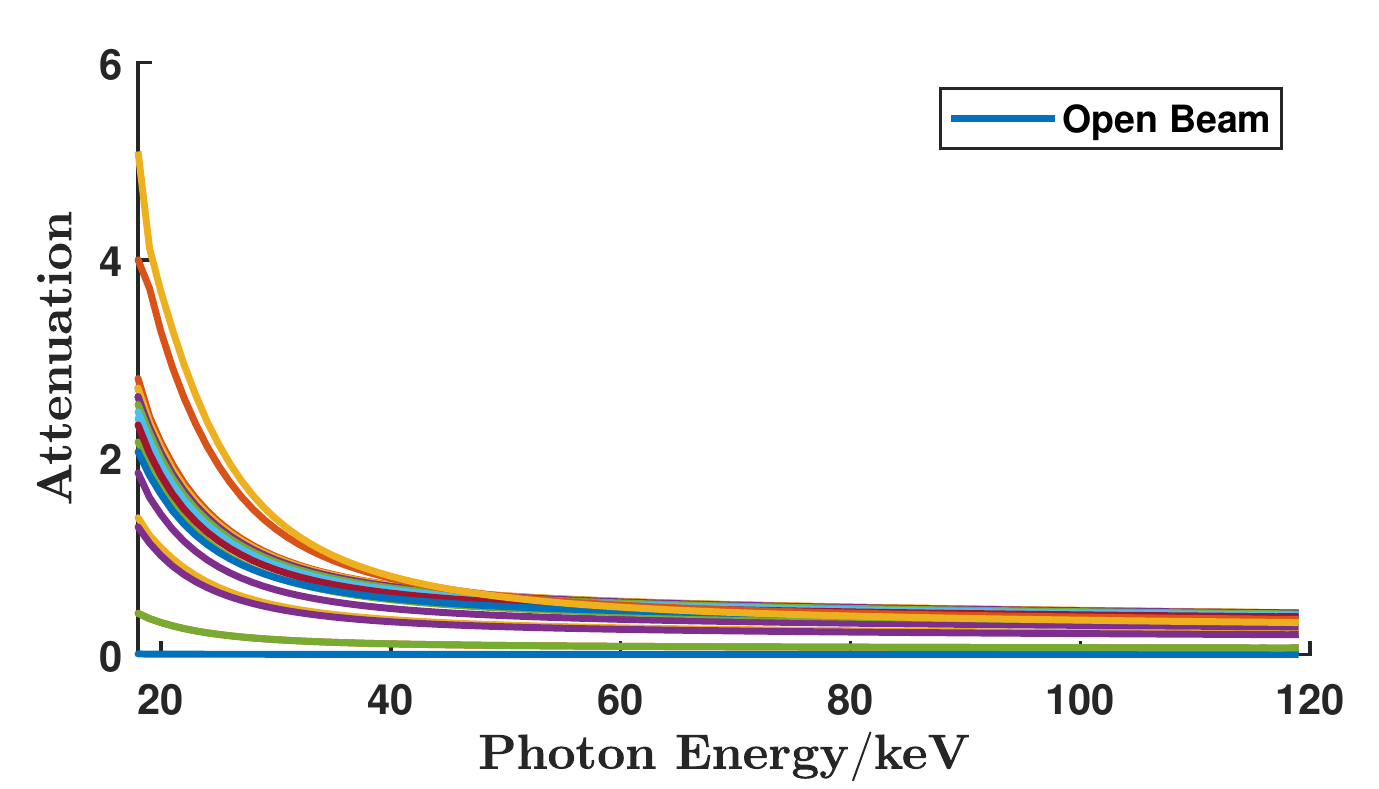}
     \subcaption{Spectral attenuation}
   \end{minipage}
   \caption{Spectrum differences of the 20 equally-spaced points on the rotation axis in the projection at the $180^\circ$ angle view, and (a) spectra of the projection at those points; and (b) the spectral attenuation after open beam correction for those points.}
   \label{fig:Spectrum_differences}
\end{figure}

With the X-ray spectrum and material phantoms, 180 spectral projections were generated for each phantom in the parallel beam configuration for convenience and without loss of generality. The projection directions are parallel to the $xoy$ plane, as illustrated in Fig. \ref{fig:PhantomG}, and the sample projections at different energies and corresponding reconstructions on one of the slices are shown in Fig. \ref{fig:PhantomRecon}. From the mono energy reconstruction slices, we can find that the attenuation of different materials changes differently with the X-ray energy. In addition, the spectra defined on 20 equally distributed points on the center $z$ axis of the projection at the view angle of $180^\circ$ are shown in Fig. \ref{fig:Spectrum_differences}. The spectra vary significantly in both shapes and magnitude due to different material compositions along the projection paths.

\subsection{Charge splitting model}\label{sec:CS}
With those generated projection data representing different spectral attenuation properties, charge splitting effects inside the PCD were simulated using the Photon Counting Toolkit (PcTK version 3.2) developed by Taguchi \cite{taguchi2016spatio,taguchi2018spatio}. As a brief introduction, the model considers different cases of major interactions between X-ray photons within the diagnostic energy range and the PCD sensor crystal (cadmium telluride, CdTe) in a probability framework, including free penetration with no detection, full detection without fluorescence, partial detection with fluorescence lost, and full detection with fluorescence reabsorbed. It is worth noting that the Compton scattering and Rayleigh scattering are neglected in the model due to the small probability and negligible impacts. Also the model assumes that one photon undergoes only one phenomenon whose effectiveness was validated in Monte Carlo simulation.  

Charge splitting happens in both detection cases with or without fluorescence generation,
where primary charge electron clouds could be split by neighboring pixels, which will shift the reported energy downward but with an increase in counts in these lower energy bins. In addition, in the case of partial detection with fluorescence loss, the primary electron clouds lose the energy the fluorescence photon carries, enhancing the low energy shifting. For the full detection with fluorescence reabsorbed, the secondary electron clouds, generated by the fluorescent photon re-absorption at a nearby location offset by a certain traveling distance from the initial X-ray photon incident position, can also be shared with neighboring pixels, which adds to the low energy tail. 

With one X-ray photon of energy $E_i$ incident on a pixel, the count expectations in all energy channels of nine pixels (with eight surrounding pixels) are recorded in the diagonal elements in the normalized covariance matrix $nCov3\text{x}3E_i$ with 1-keV-width energy windows. The electronic noise is added through the convolution with a one dimensional zero-mean Gaussian kernel characterized by a parameter $\sigma_e$ along the diagonal elements, leading to energy resolution degradation. The diagonal of the matrix is also termed as the spectral response function of the detector. Correlated Poisson noise can be generated with the help of the normalized covariance matrix.

In our case, we used the PcTK tool in the non-correlation mode since each projection generates one spectral image, and we do not care about the correlations of noises between pixels, hence we were only interested in the diagonal elements in the covariance matrix which represent the spectral response function of a detector. The Poisson noise is added pixel by pixel.

\begin{figure}
   \begin{minipage}[b]{.4\linewidth}
     \centering
     \includegraphics[width=1.0\textwidth]{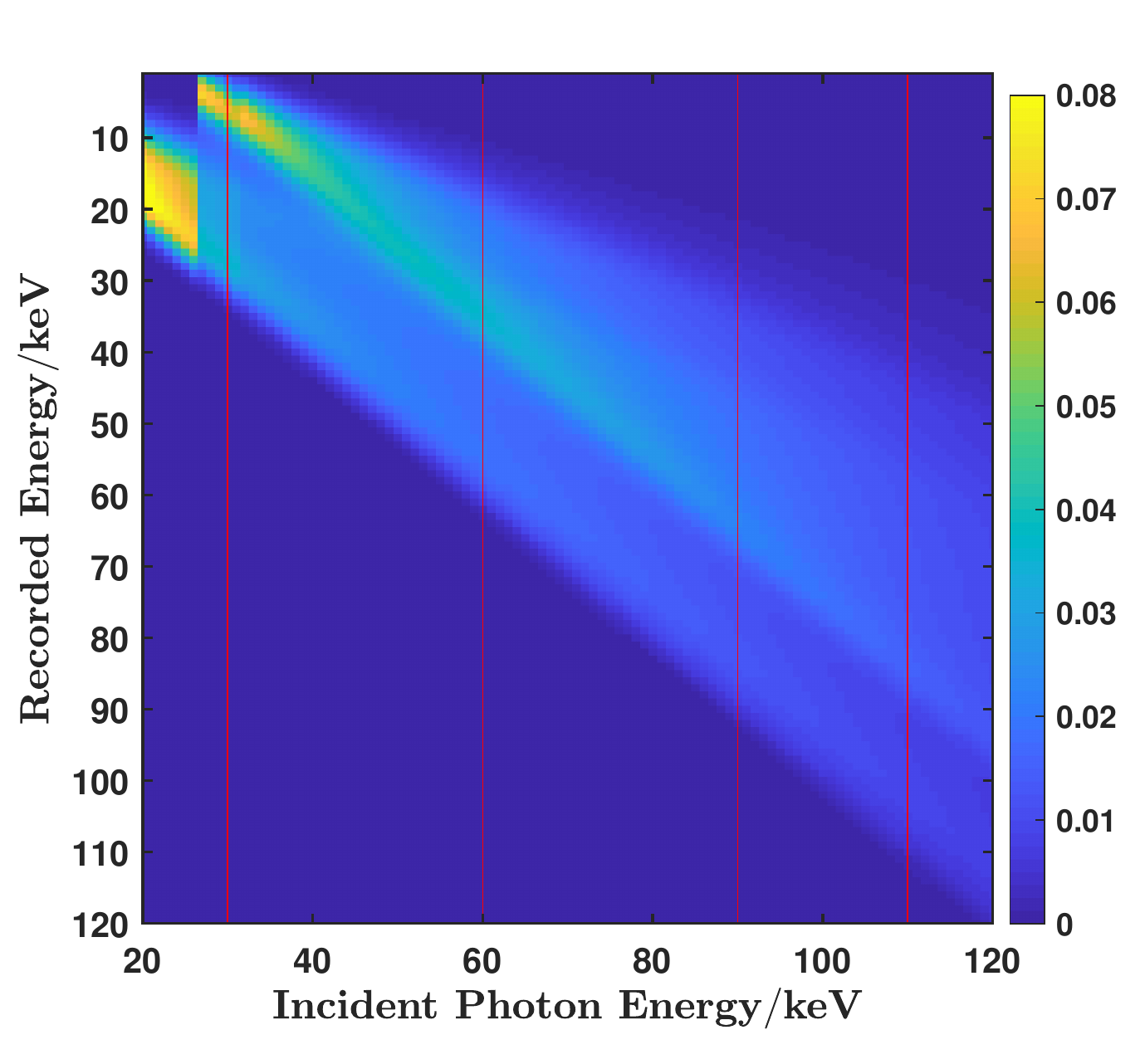}
     \subcaption{Detector spectral response}\label{fig:SRFimg}
   \end{minipage}
   \hfill
   \begin{minipage}[b]{.5\linewidth}
     \centering
     \includegraphics[width=1.0\textwidth]{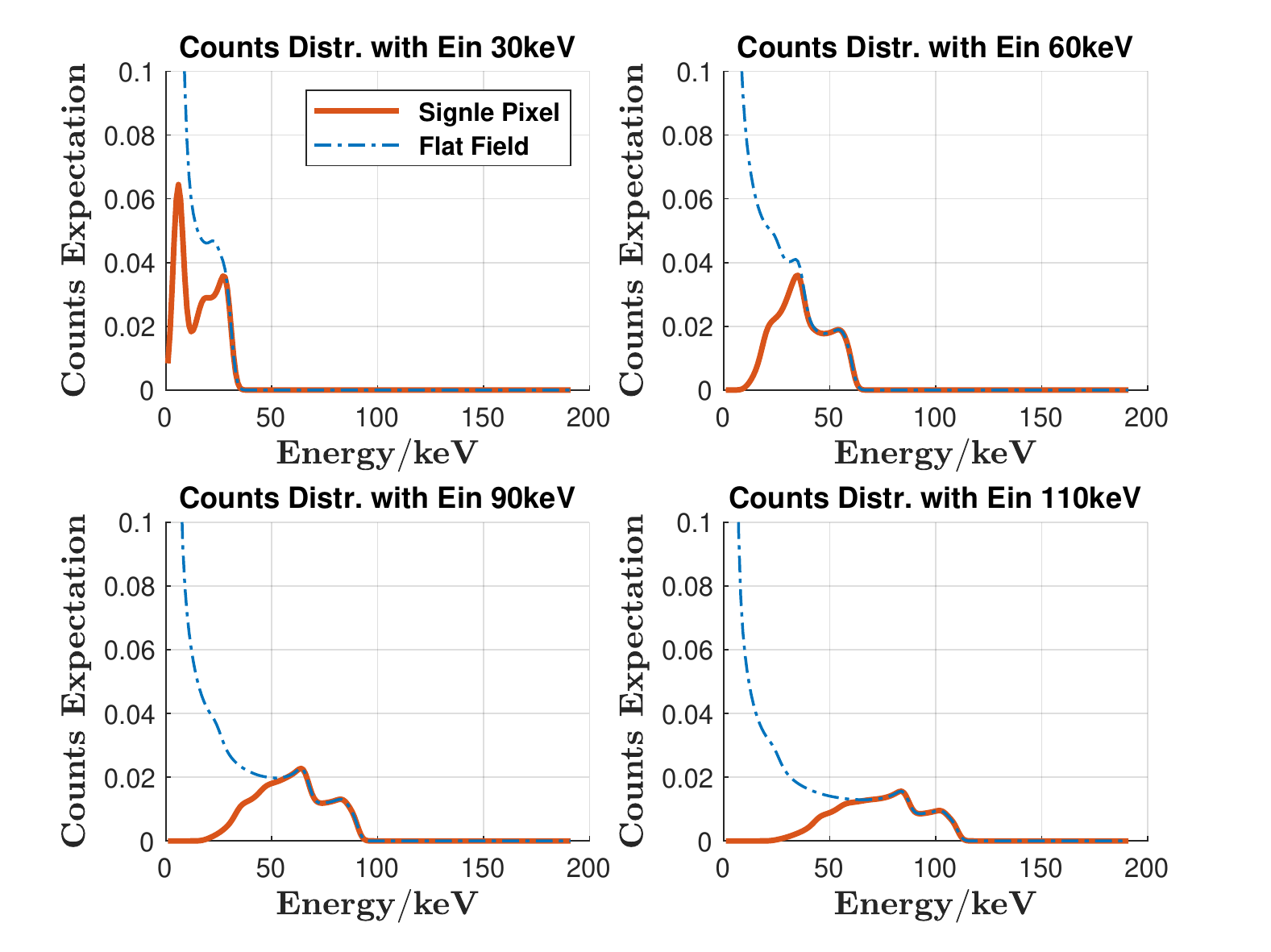}
     \subcaption{Distorted spectral responses}\label{fig:SRFcurve}
   \end{minipage}
   \begin{minipage}[b]{.9\linewidth}
     \centering
     \includegraphics[width=1.0\textwidth]{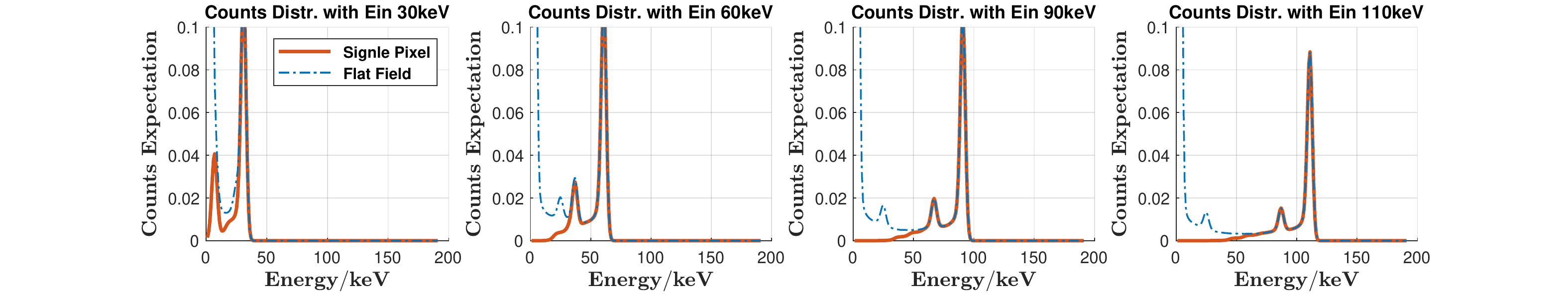}
     \subcaption{Good spectral responses with large detector pixels}\label{fig:SRFcurve_ref}
   \end{minipage}
   \caption{Spectral responses of PCDs. (a) The center pixel spectral response in the single pixel irradiation mode, and the value represents the expectation of recorded counts at energy $E_{out}$ with an incident photon of energy $E_{in}$; (b) the distorted spectral response functions [along the red vertical lines in (a)] with incident photons of $30keV$, $60keV$, $90keV$ and $110keV$ respectively; (c) Good spectral response examples that reasonably preserve spectral characteristics. The results of (a) and (b) were simulated with PcTK 3.2 with the following parameters: $r_0=24\mu m$, $\sigma_e=2.0 keV$, $d_{pix}=110\mu m$ and $d_z=2.0mm$, and $d_{pix}$ was changed to $500\mu m$ to obtain (c).}
   \label{fig:DetectorSRF}
\end{figure}

The simulation was performed with the following parameters:
\begin{inparaenum}[(1)]
\item the effective electron clouds radius $r_0$ was $24\mu m$;
\item the electronic noise parameter $\sigma_e$ was $2.0 keV$;
\item the pixel size of the detector $d_{pix}$ was $110\mu m$; and
\item the thickness of the sensor $d_z$ was $2.0mm$.
\end{inparaenum}
$r_0$ and $\sigma_e$ were selected as default which provided the best agreement with Monte Carlo simulation \cite{taguchi2018spatio}, while the other two $d_{pix}$ and $d_z$ were selected in reference to the real parameters of the Medipix-3 PCD. The center pixel spectral response function of the detector under the single pixel irradiation mode, where only the pixel of interest receives the X-rays, is shown in Fig. \ref{fig:SRFimg}. The spectral responses to several incident energy points (profiles of red lines in Fig. \ref{fig:SRFimg}) are illustrated in Fig. \ref{fig:SRFcurve} for demonstration of charge splitting effects. To be mentioned, the two energy points 27 keV and 32 keV where there are jumps in Fig. \ref{fig:SRFimg} correspond to the K-shell binding energies of cadmium (Cd) and telluride (Te) respectively. The widespread of the the recorded energy range is because of the severe charge sharing and fluorescence escaping effects due to the small pixel size, while the reasonable pixel size to preserve good spectral characteristics is at least $500\mu m$ \cite{taguchi2013vision}. In Fig. \ref{fig:SRFcurve}, the flat field mode means that all pixels receive  X-rays, and the responses include both spill-out and spill-in cross talks between pixels. It can be observed that the spectral responses are quite distorted by the charge splitting effects, since sharp peaks at the energy of incident photons have disappeared in contrast to Fig. \ref{fig:SRFcurve_ref} simulated with much larger pixels.

\begin{figure}
   \begin{minipage}[b]{.5\linewidth}
     \centering
     \begin{minipage}[b]{.49\linewidth}
        \includegraphics[width=1\textwidth]{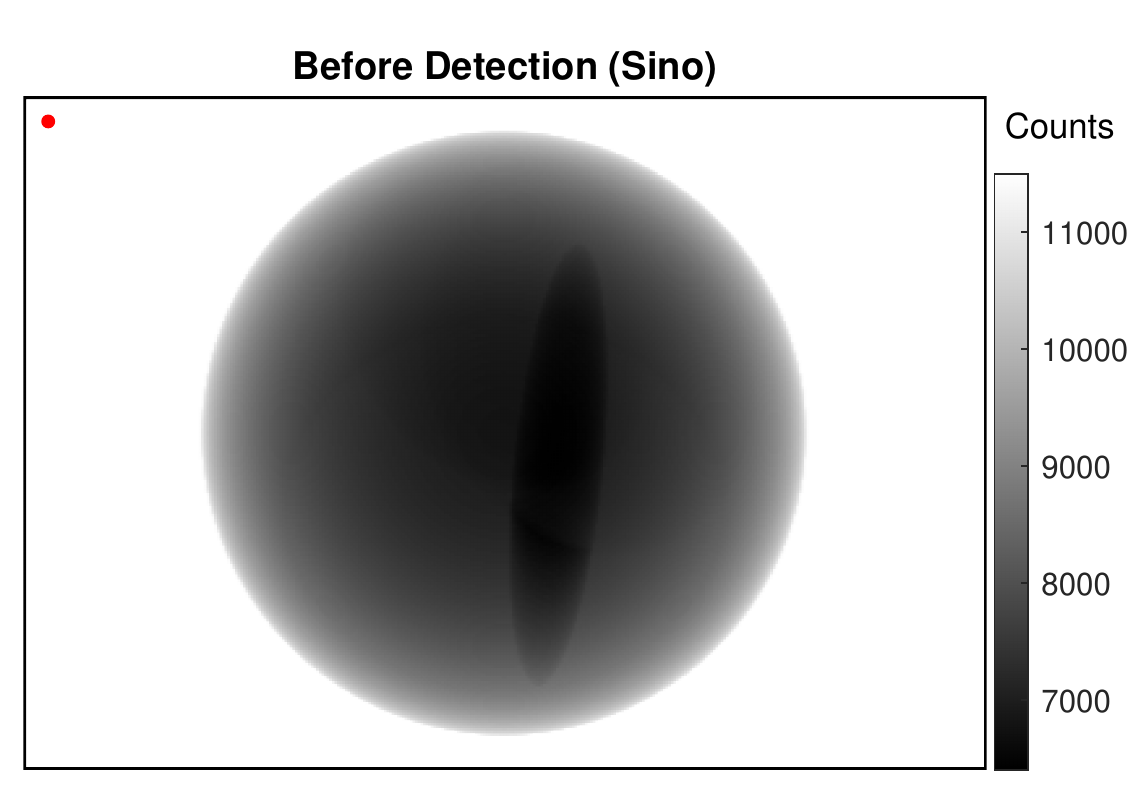}
        \includegraphics[width=1\textwidth]{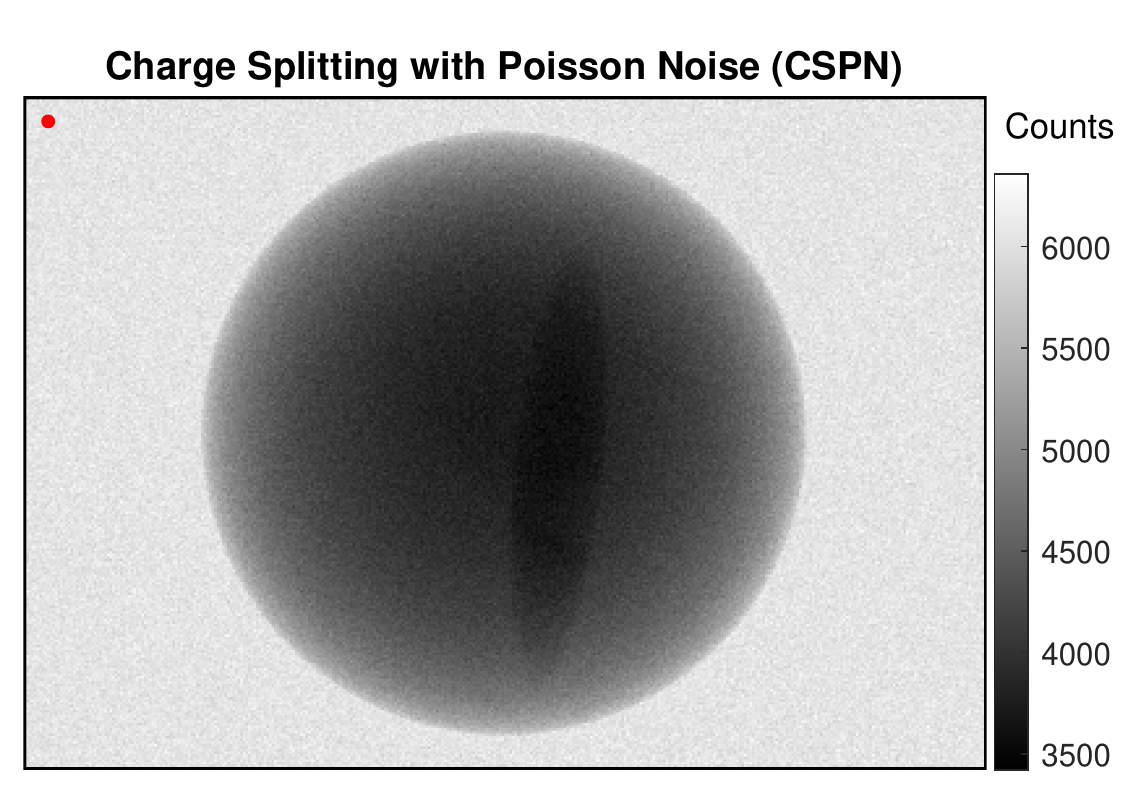}
     \end{minipage}
     \hfill
     \begin{minipage}[b]{.49\linewidth}
        \includegraphics[width=1\textwidth]{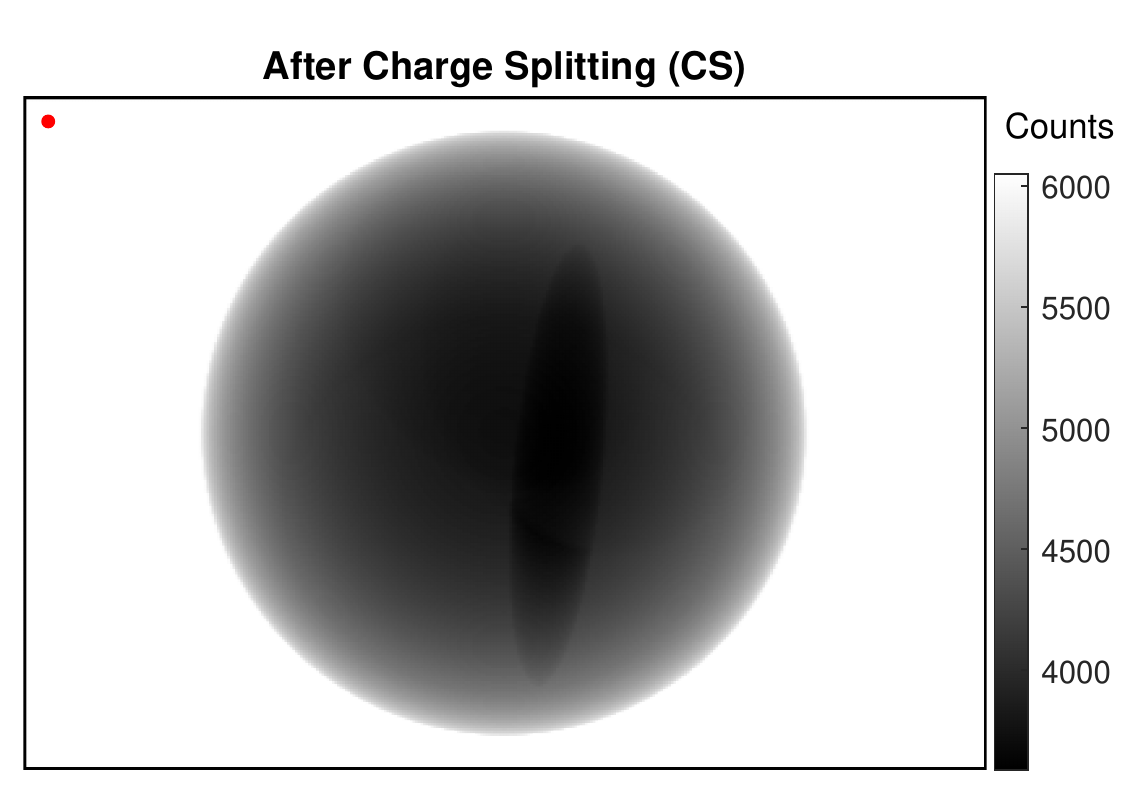}
        \includegraphics[width=1\textwidth]{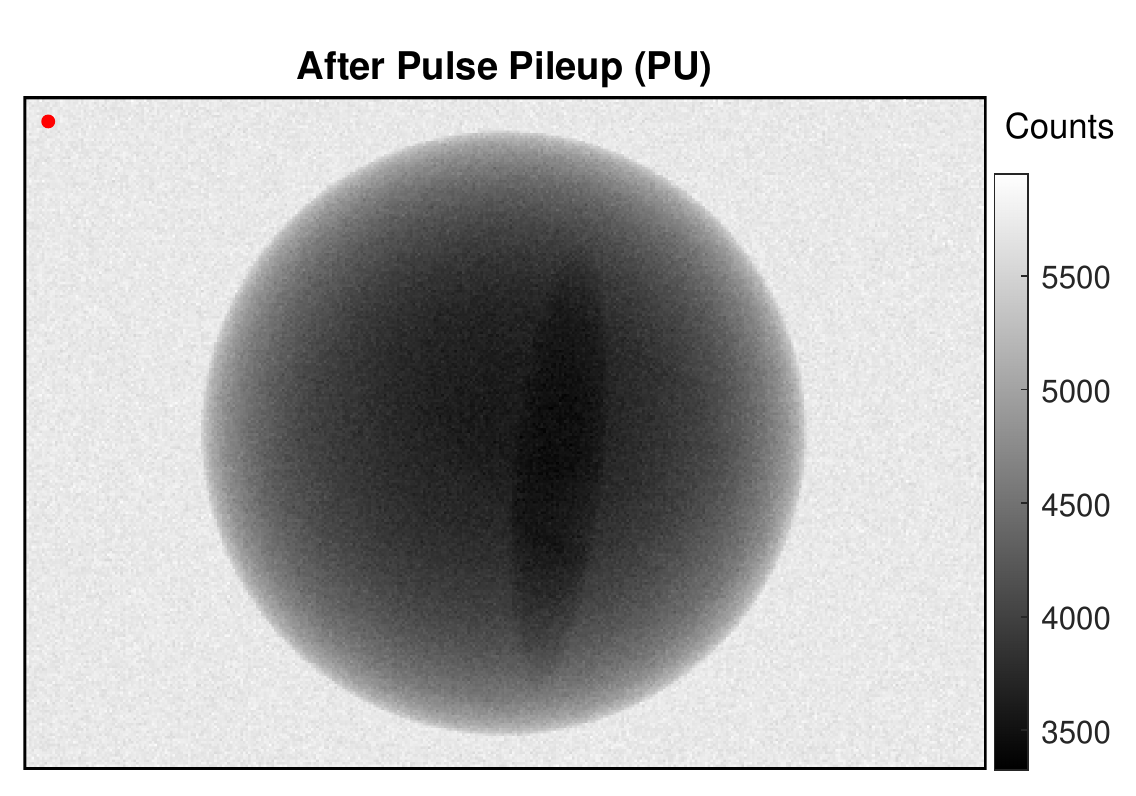}
     \end{minipage}
     \subcaption{Projections at $90^\circ$ angle view in the $60keV$ energy window}\label{fig:Projs}
   \end{minipage}
   \hfill
   \begin{minipage}[b]{.45\linewidth}
     \centering
     \includegraphics[width=1\textwidth]{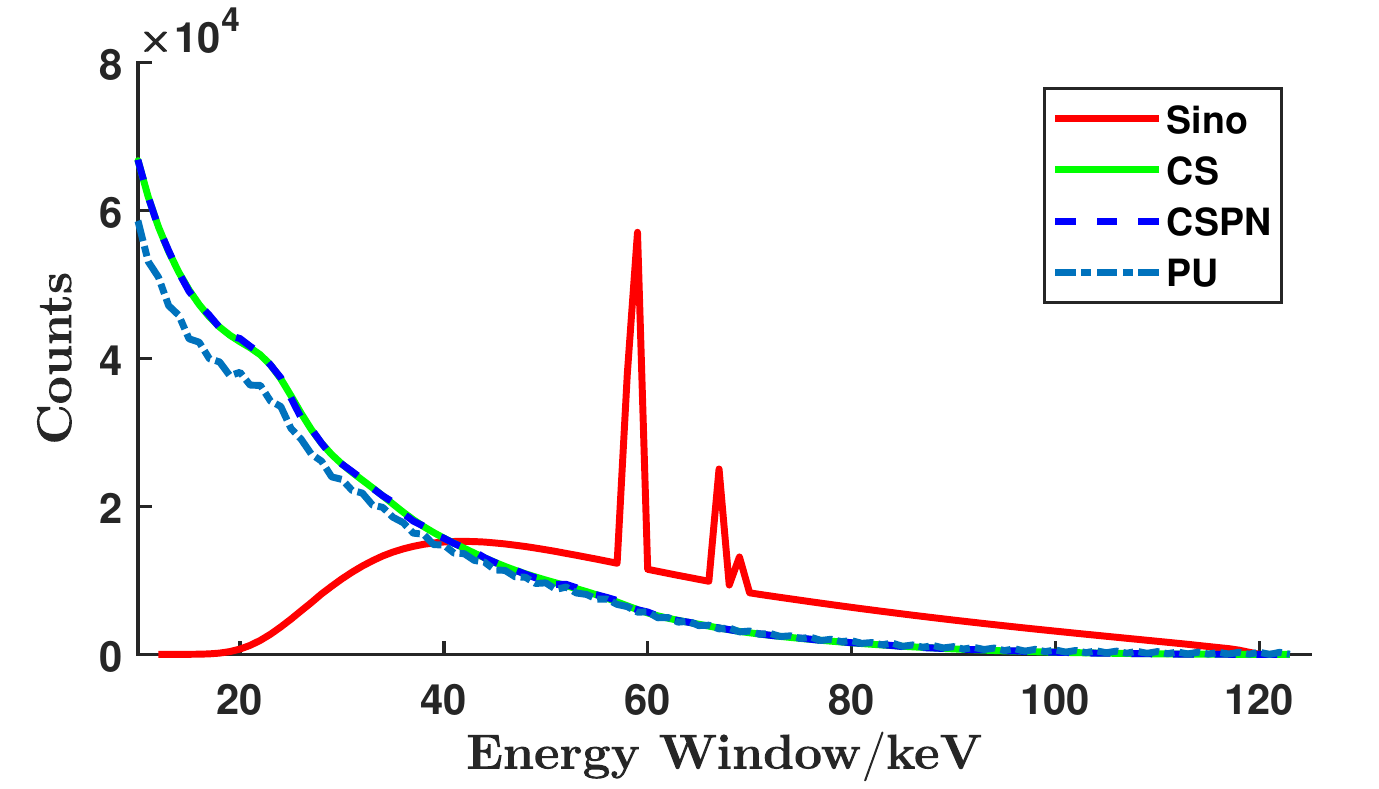}
     \subcaption{Spectral profile in an open beam area}\label{fig:SpectralProfile}
   \end{minipage}
   \caption{Projection views with charge splitting, Poisson noise and pulse pileup effects. (a) $90^\circ$ projection views in the $60keV$ energy window before detection, after charge splitting, with charge splitting and Poisson noise, and further with pulse pileup, with the gray-scale bar indicating the counts in the corresponding images; and (b) spectral counts profiles in an open beam area [indicated by the red dot near left top corner in (a)].}
   \label{fig:ProjIllustration}
\end{figure}

Combined with the pre-generated projections, the total spectral response of the detector were calculated to generate the CS-distorted spectral images. After that, Poisson noise was added to generate noisy spectral projection images. Some projection views are illustrated in Fig. \ref{fig:ProjIllustration}. It can be seen that the detected energy spectrum is distorted from that in Fig. \ref{fig:SpectralProfile}, and the distribution of counts is heavily shifted to low-energy due to the charge splitting effects. 

\subsection{Pulse pileup model}\label{subsec:PU} 
The noisy CS-distorted spectral images were then fed to the pulse pileup model to generate the final PCD detection images. The rationale is that the charge splitting effect decides the distribution of charge clouds on the pixel grids, and the pulse pileup comes from the readout process of the charge clouds. The explicit photon arrivals are determined out of randomness when the charge clouds forms. In other words, the Poisson noise should be added during the charge splitting process and the measurements will destroy the superposition states of the photons. Correlated Poisson noise is generated due to the charge splitting effects. Since we do not care about the noise correlation between pixels and each pixel itself behaves in a Poisson manner, the Poisson noise could be added after the charge splitting process to achieve equivalent observations in our cases before the pulse pile-up.

Our paralyzable model for the pulse pileup effect is based on Bierme and Roessl's work\cite{bierme2012fourier, roessl2016fourier} since PcTK does not include this effect. Modifications are made on Roessl's model to account for the spatial cross-talk between adjacent pixels receiving different spectra.
The pileup distortion is calculated in a pixel-wise manner due to its intrapixel effect nature, hence, let us focus on one pixel and study the readout process of the recorded noisy spectrum $TR(E)$ of equivalent photons received by the pixel (each part of the electron clouds split by the charge splitting effect is treated as an equivalent photon). The pile up problem can be restated as the level-crossing problem, computing the up-crossing times $N_X(U)$ of $X(t)$ over a pre-defined threshold $U$, where $X(t)$ is the analog output of a charge sensitive amplifier followed by a pulse sharper circuit in one pixel and defined as
\begin{equation}
    X(t) = \sum_{j\in \mathbf{Z}}U_j g(t-t_j),
\end{equation}
where $U_j$ value represents the signal height of photon $j$, which are independent and identically distributed (iid) variables following the distribution of the probability mass function $p(U)$ of the ``pulse height spectrum", and also independent of $t_j$; $g(t)$ is the signal shape kernel, and $t_j$ stands for the interaction time of photon $j$. Typically, the pulse heights are in the unit of `$mV$', and the value of the pulse height usually linearly corresponds to the energy of the equivalent photon. Hence we do not discriminate pulse height $U$ and energy $E$ for simplicity, and directly denote the pulse height as the response energy level in the unit of `$keV$'. Thus, the pulse height spectrum $p(U)$ is actually the normalized spectral distribution of $TR(U)$ which is the pixel's recorded total response of the incident spectral projection and has been calculated in Subsection \ref{sec:CS}. More precisely,
\begin{equation}
    \lambda = \int_{-\infty}^{\infty}{TR(U)dE}
\end{equation}
\begin{equation}\label{eq:pU}
    p(U)=\frac{TR(U)}{\lambda}
\end{equation}
where $\lambda$ represents the total count of the equivalent photons.

Based on Bierme's mathematical deductions \cite{bierme2012fourier}, for finite signal heights $U_j$, as long as the signal shape kernel $g$ is at least piece-wise second order differentiable and non-increasing on each interval, $N_X(U)$ can be calculated with Eq. \ref{eq:NXu} in the Fourier domain as: 
\begin{equation}\label{eq:NXu}
    \hat{N_X}(u)=\lambda\exp\left[\lambda\int_{-\infty}^{\infty}{\left(\hat{P}[ug(t)]-1\right)dt}\right]\sum_{tj:\Delta g(t_j)>0}\frac{\hat{P}\left[u g(t^+_j)\right]-\hat{P}\left[u g(t^-_j)\right]}{iu}. 
\end{equation}
In Eq. \ref{eq:NXu}, $\Delta g(t_j)=g(t_j^+)-g(t_j^-)$, where $g(t_j^+)$ and $g(t_j^-)$ represent the right and left limits at jumps. $\hat{P}(u)$ and $\hat{N_X}(u)$ are the inverse and forward Fourier transforms of $p(U)$ and $N_X(U)$, respectively. $i$ is the imaginary unit. The requirement is actually quite easy to meet in practice, since we can fit the rising parts of any smooth kernel shapes with piece-wise constant functions at arbitrary accuracy. With such approximation, most of right and left limit contributions to the sum over positive jumps are cancelled out.

In this study, we used normalized monopolar pulses as the kernel shape for simulation which are non-negative and with one single maximum. Thus, Eq. \ref{eq:NXu} reduces to
\begin{equation}\label{eq:NXu_sim}
    \hat{N_X}(u)=\lambda\exp\left[\lambda\int_{-\infty}^{\infty}{\left(\hat{P}[ug(t)]-1\right)dt}\right]\frac{\hat{P}(u)-1}{iu}. 
\end{equation}
We define the cumulative total response function $\Phi_{TR}(U)$ as
\begin{equation}
    \Phi_{TR}(U) = \begin{cases}
                    \int_U^\infty{TR(U^\prime)dU^\prime}, &  U\geq0\\ 
                    0, & U<0
                \end{cases}
\end{equation}
and we have the relationship between $TR(U)$ and $\Phi_{TR}(U)$ in the Fourier spectrum space as:
\begin{equation}\label{eq:Phi_u}
     \hat{\Phi_{TR}}(u) = \frac{\hat{TR}(u)-\hat{TR}(0)}{iu} 
\end{equation}
By substituting Eqs. \ref{eq:pU} and \ref{eq:Phi_u} into Eq. \ref{eq:NXu_sim}, we obtain
\begin{equation}
   \hat{N_X}(u) =\hat{\Phi_{TR}}(u)\exp{[-\hat{TK}(u)]}
\end{equation}
where 
\begin{equation}
    \hat{TK}(u)=\int_{-\infty}^{-\infty}{\big\{\hat{TR}(0)-\hat{TR}[ug(t)]\big\}dt}
\end{equation}

Finally, the recorded counts above the energy threshold $U$ with the pulse pileup effect can be calculated by transforming $\Hat{N_X}(u)$ back to the count space:
\begin{equation}
    N_X(U)=\frac{1}{2\pi}\int_{-\infty}^{-\infty}{\hat{\Phi_{TR}}(u)\exp{[-\hat{TK}(u)]}\exp{(-iuU)}}du
\end{equation}
It is worth noting that by using the total spectral response calculated in Subsection \ref{sec:CS} instead of the detector spectral response function in ref\cite{roessl2016fourier}, the spatial cross-talk between pixels due to the charge splitting effect is included in our model, while the model in the ref\cite{roessl2016fourier} only fits flat-field situations.

\begin{figure}[htbp]
  \centering
  \begin{minipage}[b]{0.45\textwidth}
    \includegraphics[width=\textwidth]{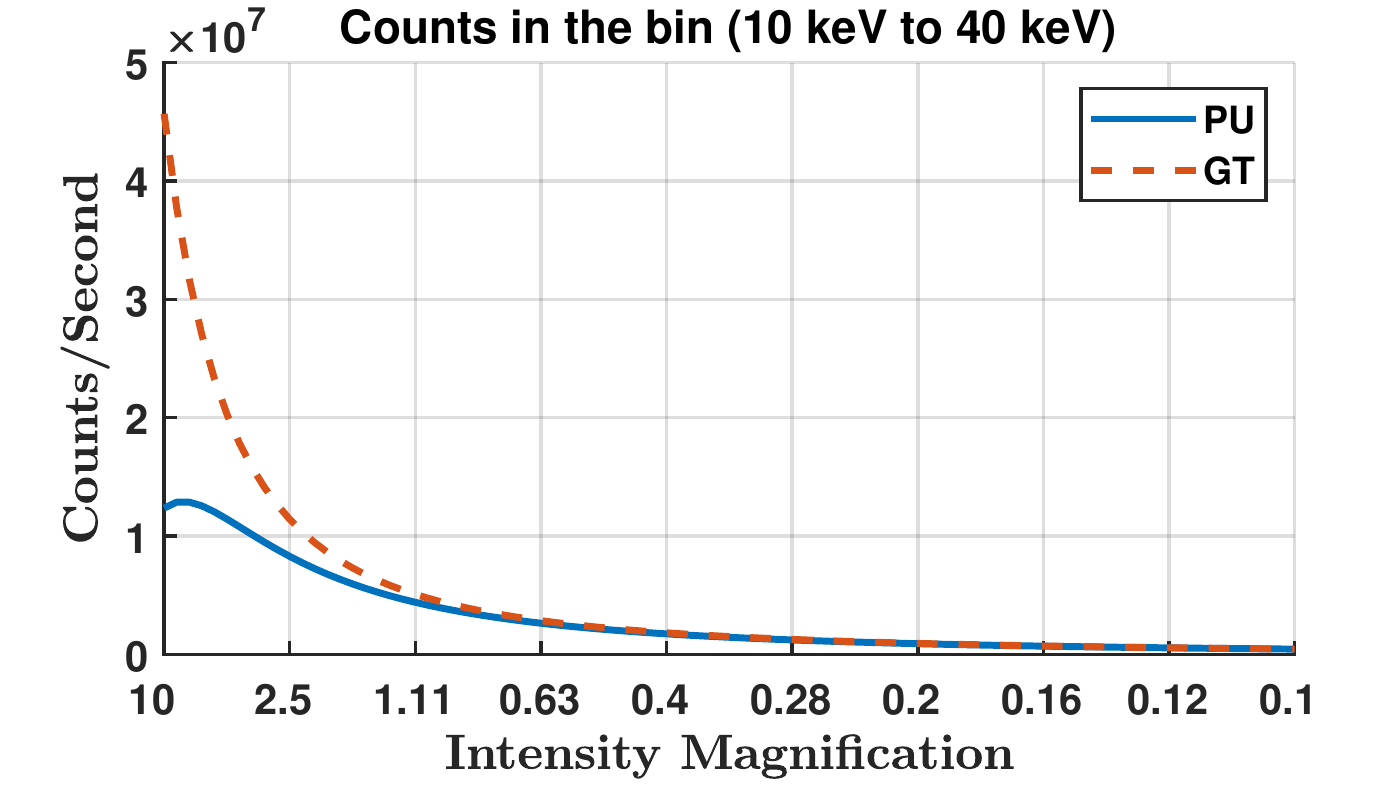}
    \caption{Pulse pileup effect illustration, where PU denotes the distorted detected counts due to the pileup effect, and GT is for the ground truth.}
    \label{fig:PUdemo}
  \end{minipage}
  \hfill
  \begin{minipage}[b]{0.45\textwidth}
    \includegraphics[width=\textwidth]{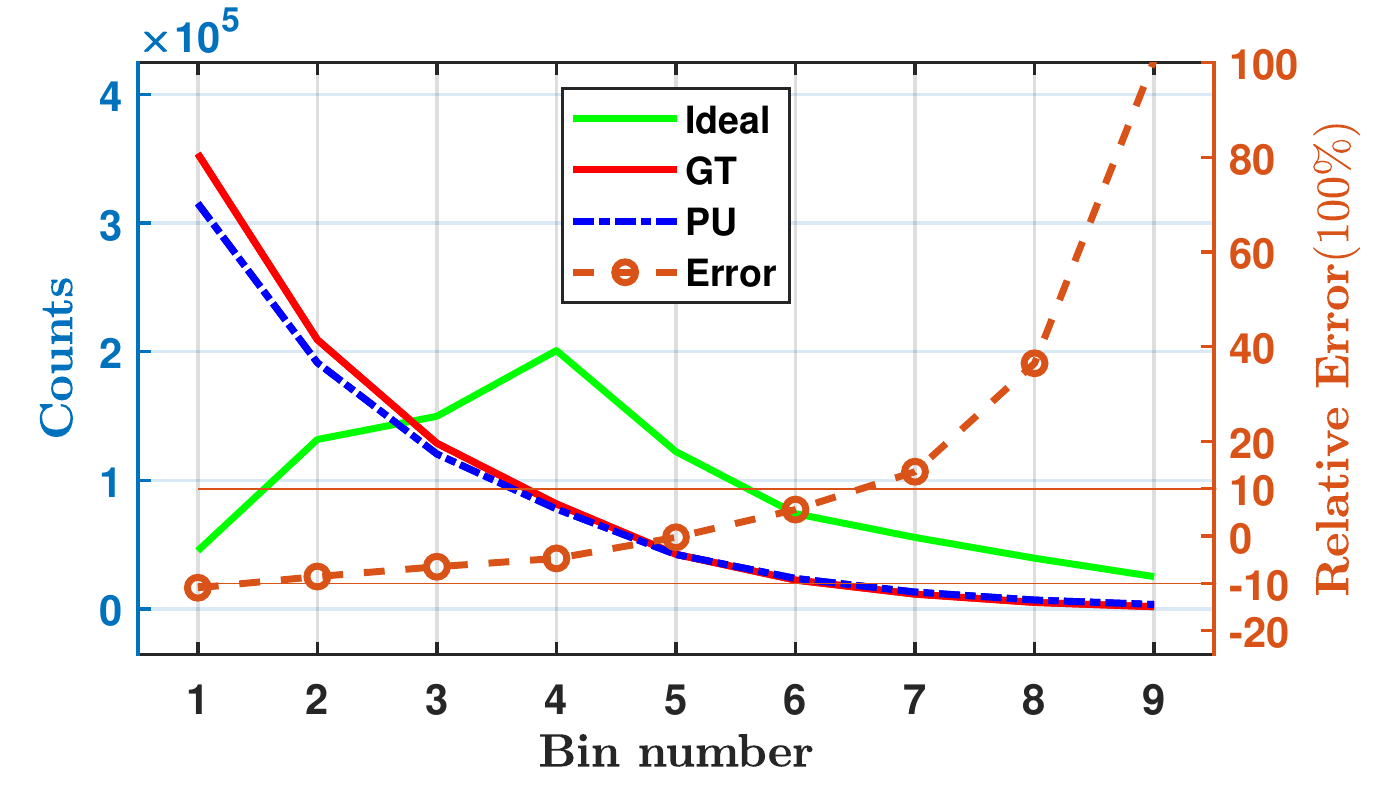}
    \caption{Detected counts in $10keV$-width energy bins from 20 to $110 keV$. Relative error was calculated as (PU-GT)/GT.}
    \label{fig:PUerrorDemo}
  \end{minipage}
\end{figure}

The pulse shapes used for data generation in our case are $g(t)=e^{-t/\tau}$ for positive $t$ and zero otherwise, and the dead time $\tau$ was set as $10ns$. To illustrate the pulse pileup effect, the X-ray intensity was first raised to ten times the intensity setting for our data generation as demonstrated in Fig. \ref{fig:ProjIllustration}, and then gradually reduced to one tenth of the initial value. Here the ground truth refers to the detected counts after the charge splitting effect but before the pulse pileup effect. The detected counts per second during the process was illustrated in Fig. \ref{fig:PUdemo}. In the figure, the detected counts first increase and then decrease as the real photon counts monotonically decrease, which is a typical phenomenon caused by the pulse pileup effect. The detected counts are always smaller than the ground truth value because the bin detects the lower energy part of the spectrum (Fig. \ref{fig:SpectralProfile}) and it tends to lose counts when pulse pileup events happen. If the bin detects the range in the middle part of the spectrum, the situation will be more complicated since the curve tendency will be a balance between the count increase due to intensity rise and lower energy photons pileup into the energy window, and the count loss due to the pileup of photons within the energy window. 

Since the spectral resolution of the current PCD is not better than $5keV$, it is reasonable to choose thresholds with a step size much larger than $1keV$ to reduce the amount of involved data. With our choice of $10keV$ energy width, the thresholds used are $20keV$, $30keV$, $40keV$, $50keV$, $60keV$, $70keV$, $80keV$, $90keV$, $100keV$ and $110keV$. Based on our data generation settings, the open beam data counts in nine energy bins are shown in Fig. \ref{fig:PUerrorDemo}, where the nine bins correspond to the energy ranges $20-29keV$, $30-39keV$, ..., $100-109keV$, respectively. It is shown by the relative errors curves in Fig. \ref{fig:PUerrorDemo} that the counting errors in bins 2 to 6 are within $\pm 10\%$, bin 1 error is slightly below $-10\%$ while bin 7 error is slightly above $10\%$, and errors in bin 8 and 9 are well beyond $10\%$. Clearly, the detected signals suffer from the pulse pileup effect which shifts the counts in lower energy bins to higher energy bins. The curve \textit{Ideal} refers to the counts detected by the ideal PCD. Compared with the ideal data, it can be seen that the shape of the detected counts is signficantly distorted by the charge splitting and pulse pileup effects. Our goal is to recover the ideal curves from the distorted signals with deep learning networks.

\subsection{Correction Networks}
\subsubsection{WGAN}
A Wasserstein Generative Adversarial Network (WGAN)\cite{arjovsky2017wasserstein} was designed for this work, as shown in Fig. \ref{fig:Networks}. The goal is to build a mapping $\Ggan$ that transforms degraded projection measurements $\mb\in \mathbb{R}^{N\times N\times N_{E}}$ to the ideal projection data $\pb\in \mathbb{R}^{N\times N\times N_{E}}$. In other words, we would modify the distribution of degraded spectral projections to be as close as possible to the distribution of ideal data by refining $\Ggan$ in a data-driven manner. In the Generative Adversarial Network (GAN)\cite{goodfellow2014generative} framework, a discriminator network $\Dgan$ is introduced to help train the mapping network also referred to as the generator $\Ggan$. The generator $\Ggan$ takes the input and transforms it toward the corresponding target, while the discriminator tries to discriminate between the output of $\Ggan$ and the real target. During the training, $\Ggan$ receives feedback from $\Dgan$ and other metrics then generates more realistic results, while $\Dgan$ receives feedback from the labels for the output of $\Ggan$ and the real target then continuously improves its discrimination ability. Through this adversarial competition between $\Dgan$ and $\Ggan$, $\Ggan$ is expected to learn the characteristics of the target data distribution without involving an explicit loss function (Actually, the discriminator $\Dgan$ plays the role of the loss function guiding the training of $\Ggan$). As a result, $\Ggan$ can learn extremely complex representations of underlying data through adversarial learning. The expected final equilibrium is that the output distribution of $\Ggan$ is so close to the ground truth distribution that $\Dgan$ fails to tell differences between the two. 

As a major improvement to the generic GAN, instead of using the Jensen-Shannon divergence WGAN utilizes the Wasserstein distance to measure the difference between data distributions. The Wasserstein distance computes the cost of mass transportation from one distribution to the other. Thus, the solution space of $\Ggan$ is greatly compressed, and the training process facilitated. The adversarial loss function is expressed as follows:
\begin{equation}
    \min_{\Dgan}\max_{\Ggan} L_{WGAN}(\Dgan,\Ggan) = -\mathbb{E}_{\pb}[\Dgan(\pb)] + \mathbb{E}_{\mb}\left[\Dgan\left(\Ggan(\mb)\right)\right] + \lambda\mathbb{E}_{\hat{\pb}}\left[\left(\norm{\nabla_{\hat{\pb}}\Dgan(\hat{\pb})}_2-1\right)^2\right],
    \label{eq:Loss_wgan}
\end{equation}
where the first two terms represent the Wasserstein distance estimation, the last term is the gradient penalty which is an alternative of weights clipping to enforce the Lipschitz constraint on the discriminator for better stability\cite{gulrajani2017improved}, $\hat{\pb}$ is uniformly sampled along the straight line between paired $\Ggan(\mb)$ and $\pb$, and the penalty coefficient $\lambda$ is a weighting constant.

\subsubsection{Network structures}
Our overall network consists of one generator and one discriminator, which are mainly constructed with convolutional layers. As a reversal of the forward model, we divide our correction scheme into two steps: first to conduct the pulse pileup correction with a representation network, then to accomplish the charge splitting correction and denoising using a convolutional network. Following the idea, the generator comprises two sub-networks, dePUnet and deCSnet corresponding to the two steps respectively. 

\begin{figure}
   \begin{minipage}[b]{.49\linewidth}
     \centering
     \includegraphics[width=0.8\textwidth]{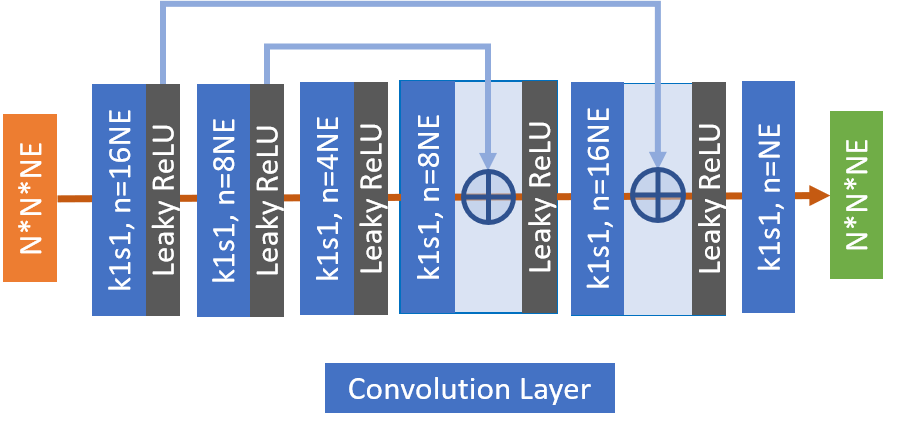}
     \subcaption{First part of the generator network dePUnet}\label{fig:dePU}
   \end{minipage}
   \hfill
   \begin{minipage}[b]{.49\linewidth}
     \centering
     \includegraphics[width=0.55\textwidth]{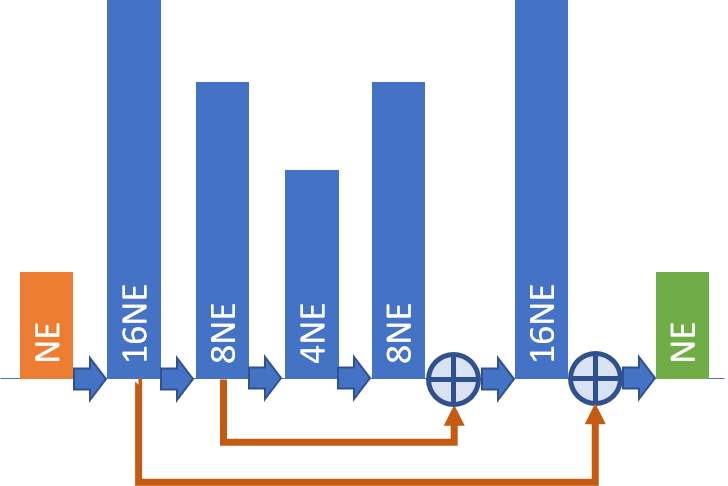}
     \subcaption{Equivalent structure}\label{fig:dePUeq}
   \end{minipage}
   \begin{minipage}[b]{.49\linewidth}
     \centering
     \includegraphics[width=0.8\textwidth]{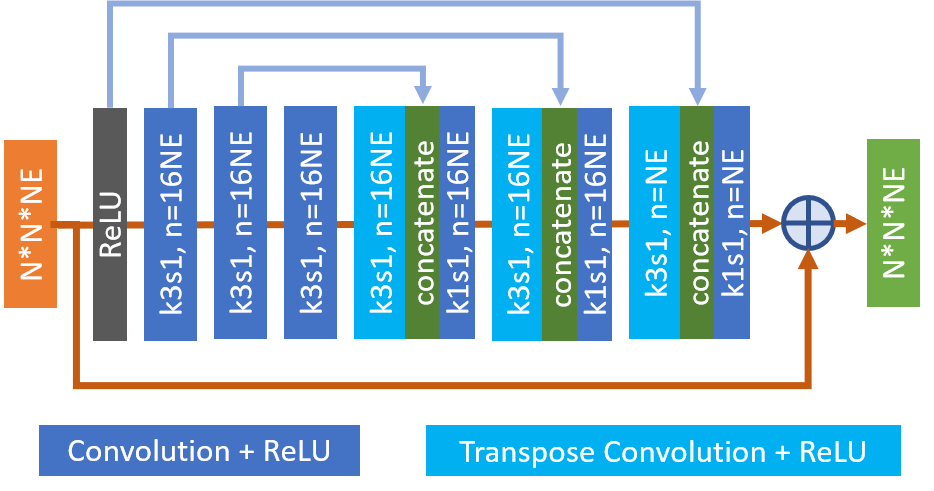}
     \subcaption{Second part of the generator network deCSnet}\label{fig:deCSnet}
   \end{minipage}
   \hfill
   \begin{minipage}[b]{.49\linewidth}
     \centering
     \includegraphics[width=0.95\textwidth]{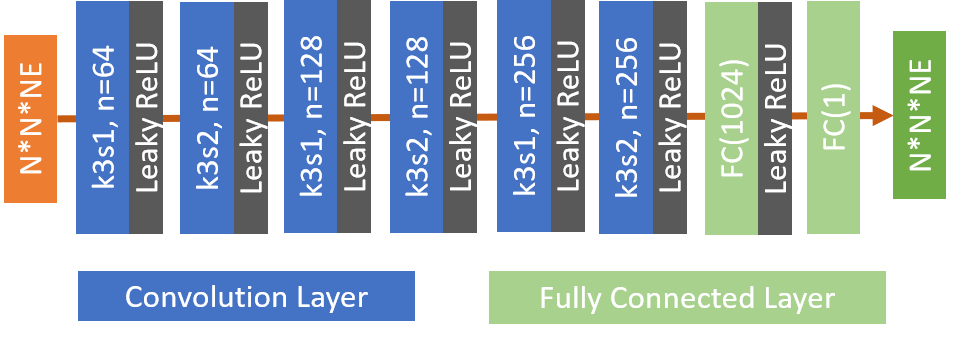}
     \subcaption{Discriminator network}\label{fig:Dgan}
   \end{minipage}
   \caption{Network structures, where $k$, $s$ and $n$ stand for kernel size, stride, and number of kernels respectively, $N_{E}$ represents the number of X-ray spectral channels (i.e., the number of energy bins, which is 9 in this study).}
   \label{fig:Networks}
\end{figure}

The structure of dePUnet is designed based on the intra-pixel effect nature of the pulse pileup, which yields cross-talk only between spectral channels of one pixel during the readout. It is built with convolutional layers of $1\times1$ kernel size as shown in Fig. \ref{fig:dePU}, and only conducts spectral transformation for each pixel. In addition, instead of correcting the noisy pileup signals to the noisy charge splitting signals, we directly correct them to the clean signals just after the charge splitting effect since it is easier for the network to learn denoising rather than to represent noisy signals due to the inherent regularization property of the network. The latter mapping strategy also makes the subsequent correction steps easier. The dePUnet is designed as an auto-encoder with shortcuts as demonstrated in Fig. \ref{fig:dePUeq}, because of the strong representation and noise suppression ability of this light-weight topology, and the shortcuts are used to ease training. The leaky ReLU activation is used for all convolutional layers except for the last layer whose output will be the clean charge splitting signals.

The deCSnet is a fully convolutional network with shortcuts and expected to achieve deconvolution and denoising, due to the fact that the charge splitting process can be expressed as a convolution operation while the deconvolution and denoising are the strengths of the fully convolultional network. We designed the deCSnet in reference to several state-of-art denoising GAN networks (WGAN-VGG\cite{yang2017low}, WGAN-CPCE\cite{shan20183d} and GAN-CNN\cite{wolterink2017generative}). Totally, we have three convolutional layers with kernel size of $3\times3$ and three corresponding transpose convolution blocks to compensate for the dimension reduction, and a ReLU layer at the beginning to connect the dePUnet output as shown in Fig. \ref{fig:deCSnet}. Inside each transpose convolution block, the output of a transpose convolution layer fed with the block input is concatenated with the intermediate result of the same dimension from previous layers outside the block, and followed by a projecting convolution layer with kernel size of $1\times1$ to reduce the channel dimension and improve the computational efficiency, and form the block output. All convolution layers and transpose convolution blocks share the same number of kernels $16 N_{E}$ ($N_{E}$ is the number of energy bins) except for the last block which has $N_E$ kernels to match the input dimension. An additional shortcut directly connects the input to the output, making the network a residual type, which is more advantageous than the direct mapping from the input to the ground truth\cite{he2016deep}. 

The data correction is undertook by the generator $\Ggan$ (dePUnet and deCSnet). Inspired by the impressive low-dose CT (LDCT) denoising results with WGAN-VGG\cite{yang2017low} and WGAN-CPCE\cite{shan20183d}, we trained the generator in the WGAN framework. Due to the similarity between tasks for the discriminators in LDCT denoising and PCD data correction, we used the same discriminator structure as those in WGAN-VGG and WGAN-CPCE. The discriminator consists of six convolution layers with the identical kernel size of $3\times3$ and two fully connection layers, and the leaky ReLU activation function is used for all layers as illustrated in Fig.~\ref{fig:Dgan}. The numbers of kernels in the convolution layers are 64, 64, 128, 128, 256, and 256 respectively, while the strides are 1 for odd layers and 2 for even layers. The output of the final convolution layer is flattened and connected to the two fully-connected layers with sizes of 1024 and 1 respectively.

\subsubsection{Loss functions}
As described in Subsections \ref{sec:CS} and \ref{subsec:PU}, the forward degradation model can be expressed as
\begin{equation}
    \boldsymbol{q}(x,y,E) = \int{\pb(x,y,E^\prime)\underset{x,y}{\otimes}\mathbf{srf}(x,y,{E}^\prime,E)\dint{E}^\prime},
\end{equation}
\begin{equation}
    \mb = f_{PU}\left[\mathbf{P}(\boldsymbol{q})\right],
\end{equation}
where $\boldsymbol{q}\in \mathbb{R}^{N\times N\times N_{E}}$ is the corresponding clean charge splitting signal of the $\mb$ and $\pb$ pair; ${\otimes}_{(x,y)}$ means the convolution operation on width and height dimensions; $\mathbf{srf}(x,y,E_{in},E_{out})$ is the spectral response function of the detector;  $\mathbf{P(\cdot)}$ stands for Poisson distribution; and $f_{PU}(E)$ represents the pulse pileup transformation. Since the combination of dePUnet and deCSnet is relatively deep and not easy to train, the clean charge splitting signal data are also fed into the network as the reference for the output of dePUnet, making the intermediate results physically meaningful and facilitate convergence.

The loss we used for the generator consists of correction error, intermediate error and generation error. The correction error measures the difference between the network output and the ground truth, and we care about both the relative error (for the open beam correction before reconstruction) and the absolute error (to improve reconstruction accuracy and avoid overweighting small values). Thus, the correction error is designed as
\begin{equation}
    L_{Correction} = \mathbb{E}_{\mb,\pb}\left|{\Ggan(\mb)-\pb}\right|^2 + \mathbb{E}_{\mb,\pb}\left|\frac{\Ggan(\mb)-\pb}{\pb+\epsilon}\right|,
\end{equation}
where the first term is the mean square error (MSE) of the absolute difference, while the second term is mean absolute error (MAE) of the relative difference. The MSE metric focuses on reducing large errors, while the MAE term of relative errors puts more penalty on discrepancies from small ground truth values, with constant $\epsilon$ in the second term set to $1\times10^{-4}$ to stabilize the ratio.

The intermediate error measures the difference between the intermediate output of dePUnet and the clean charge splitting signal, and this constrain makes the intermediate output interpretable as pileup correction results in a physical sense. The main goal of this loss is to introduce latent features and help the network training under physics-based guidance.
\begin{equation}
    L_{Guidance} = \mathbb{E}_{\mb,\boldsymbol{q}}\left|{\tilde\mb-\boldsymbol{q}}\right|^2,
\end{equation}
where $\tilde\mb$ represents the output from dePUnet.

The generation error comes from the adversarial training as indicated by Eq. \ref{eq:Loss_wgan}. 

Taking all these error terms into account, the total generator loss can be written as
\begin{equation}
    L_{\Ggan} = - \mathbb{E}_{\mb}\left[\Dgan\left(\Ggan(\mb)\right)\right] + \lambda_1 \mathbb{E}_{\mb,\pb}\left|{\Ggan(\mb)-\pb}\right|^2 + \lambda_2 \mathbb{E}_{\mb,\pb}\left|\frac{\Ggan(\mb)-\pb}{\pb+\epsilon}\right| + \lambda_3 \mathbb{E}_{\mb,\boldsymbol{q}}\norm{\tilde\mb-\boldsymbol{q}}_2 , 
\end{equation}
where $\lambda_1$, $\lambda_2$ and $\lambda_3$ are the constant balancing weights. The discriminator loss solely comes from the adversarial training objective in Eq. \ref{eq:Loss_wgan}.

\section{EXPERIMENTS}
\subsection{Experiments}
\subsection{Experimental Datasets}
We first generated 10 sets of PCD data of 10 randomly generated 3D material phantoms following the steps in section \ref{sec:Method}, and each dataset contain 180 spectral projections (each projection is of size 256 x 367 x 125) from different rotation angle views. Then based on the 10 thresholds selected in subsection \ref{subsec:PU}, counts-above-thresholds data was transformed to counts-in-bins data, and the channel size was reduced to 9 from 125 dimensions. To be mentioned, the projections before PCD detection $\pb$, after charge splitting (before Poisson noise addition) $\boldsymbol{q}$, and after the pulse pileup $\mb$ were recorded , and using as labels (ground truth) and corresponding training inputs. No open beam data were generated, instead, the air region in the outer parts of the projections were used for air correction. Those 10 PCD datasets (180x256x367x9) were randomly extracted into patches from the height and width dimensions to feed the network during the training to avoid GPU memory explosion, and each patch is of size  16 x 16 x 9. To be mentioned, despite the benefit of memory occupation, patch training also lessens the requirement on training data size since the network processing subject becomes a patch instead of a full-frame image volume which also means the equivalent data amount is greatly enlarged, and in addition, patch training will force the network to focus on a small scope which is perfect for PCD data correction because the cross-talk only happens between neighboring pixels. The extracted 1,841,400 pairs of patches were shuffled to remove the connections between pairs before feeding to the networks, and among them 92,070 pairs were used for validation. Also, we generated another 5 material phantoms and corresponding PCD datasets for testing.
\subsection{Network Training}
\begin{figure}
   \begin{minipage}[b]{.32\linewidth}
     \centering
     \includegraphics[width=0.8\textwidth]{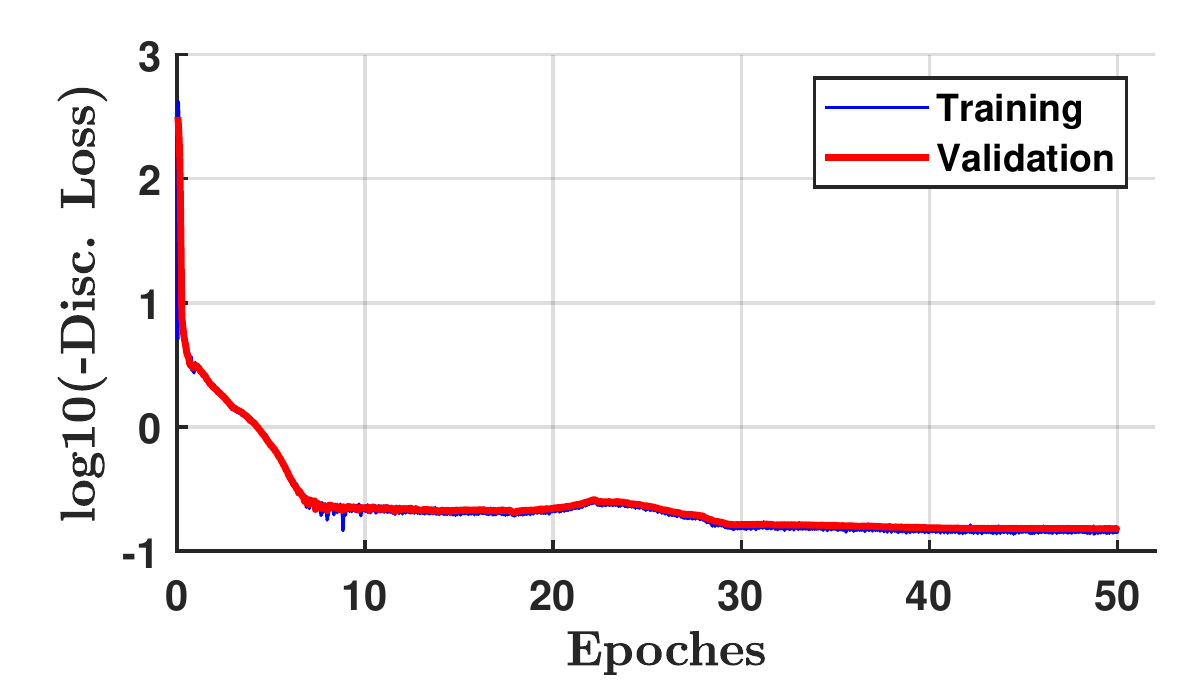}
     \subcaption{Discriminator loss}\label{fig:DistLoss}
   \end{minipage}
   \hfill
   \begin{minipage}[b]{.32\linewidth}
     \centering
     \includegraphics[width=0.9\textwidth]{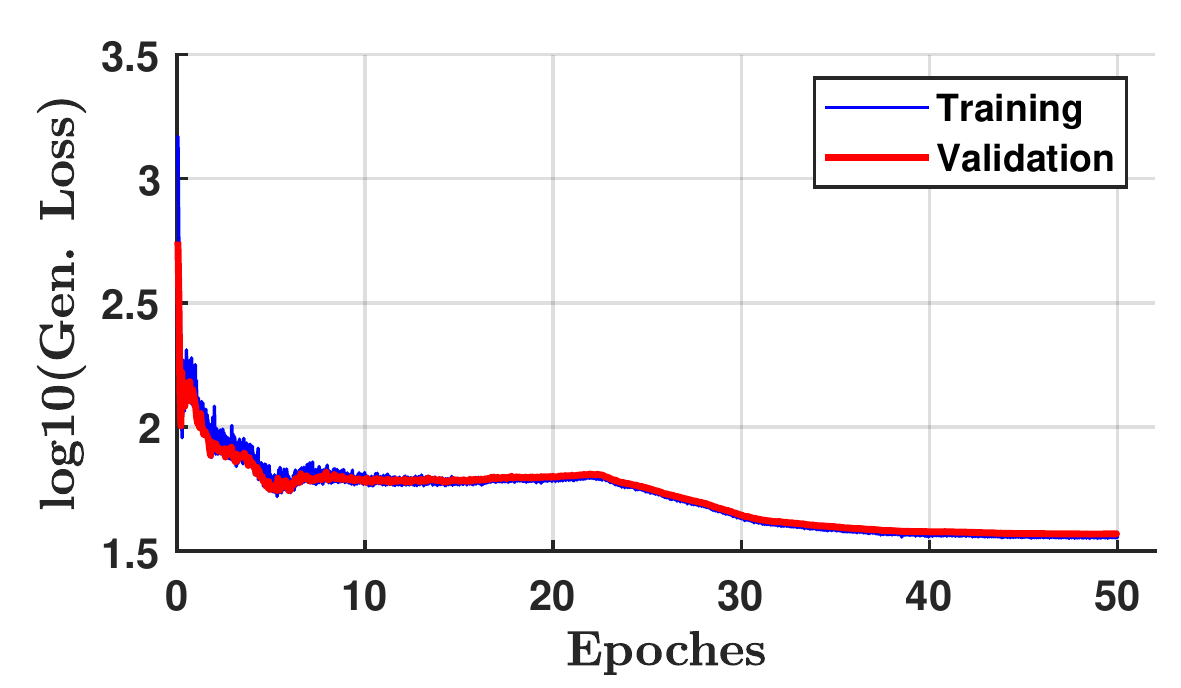}
     \subcaption{Generator loss}\label{fig:GenLoss}
   \end{minipage}
   \hfill
   \begin{minipage}[b]{.32\linewidth}
     \centering
     \includegraphics[width=0.9\textwidth]{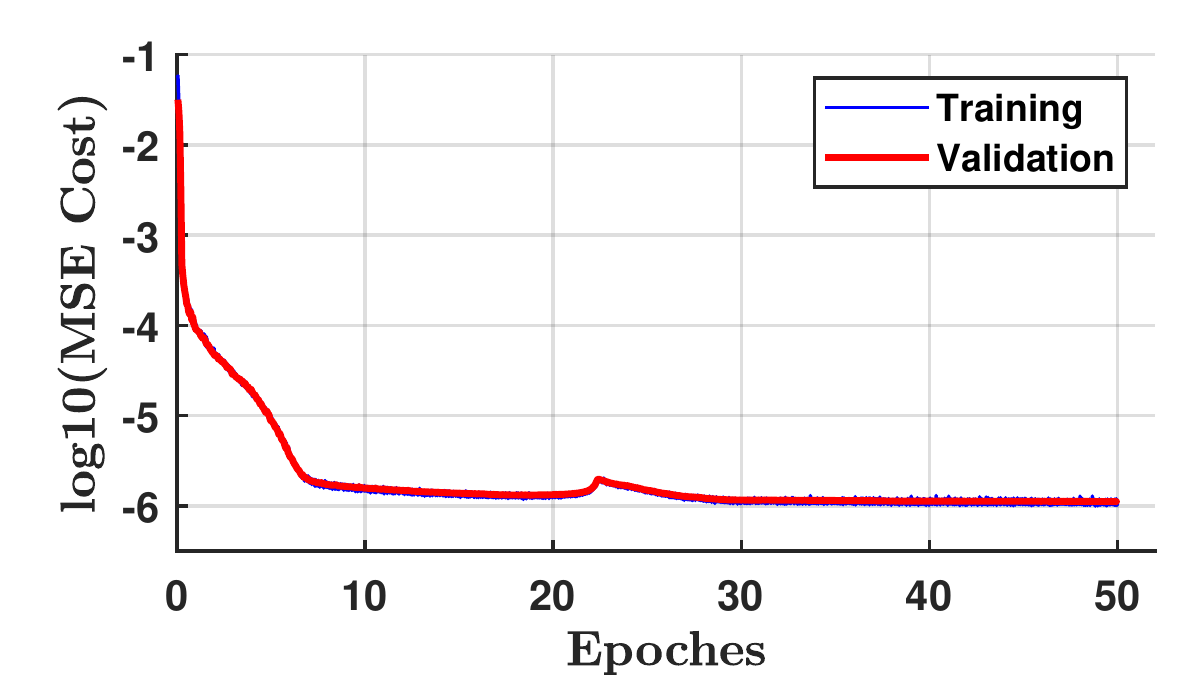}
     \subcaption{MSE cost}\label{fig:MSE}
   \end{minipage}
    \caption{Plots of discriminator loss, generator loss, and MSE cost versus the epoch number during training}
    \label{fig:training}
\end{figure}

We used Adam optimization\cite{kingma2014adam} to optimize the network with training parameters set as $\alpha = 1.0\times 10^{-4}$, $\beta_1 = 0.9$ and $\beta_2=0.999$. The mini-batch size was 1024. The penalty coefficient $\lambda$ from adversarial loss was set to 10 following the suggestion in reference \cite{gulrajani2017improved}. Hyper-parameters $\lambda_1$, $\lambda_2$, and $\lambda_3$ were all chosen as 1000. Training curves shown in Fig. \ref{fig:training} demonstrate the convergence of the network after 40 epochs.

\subsection{Correction results}
\begin{figure}
    \centering
    \includegraphics[width=0.8\textwidth]{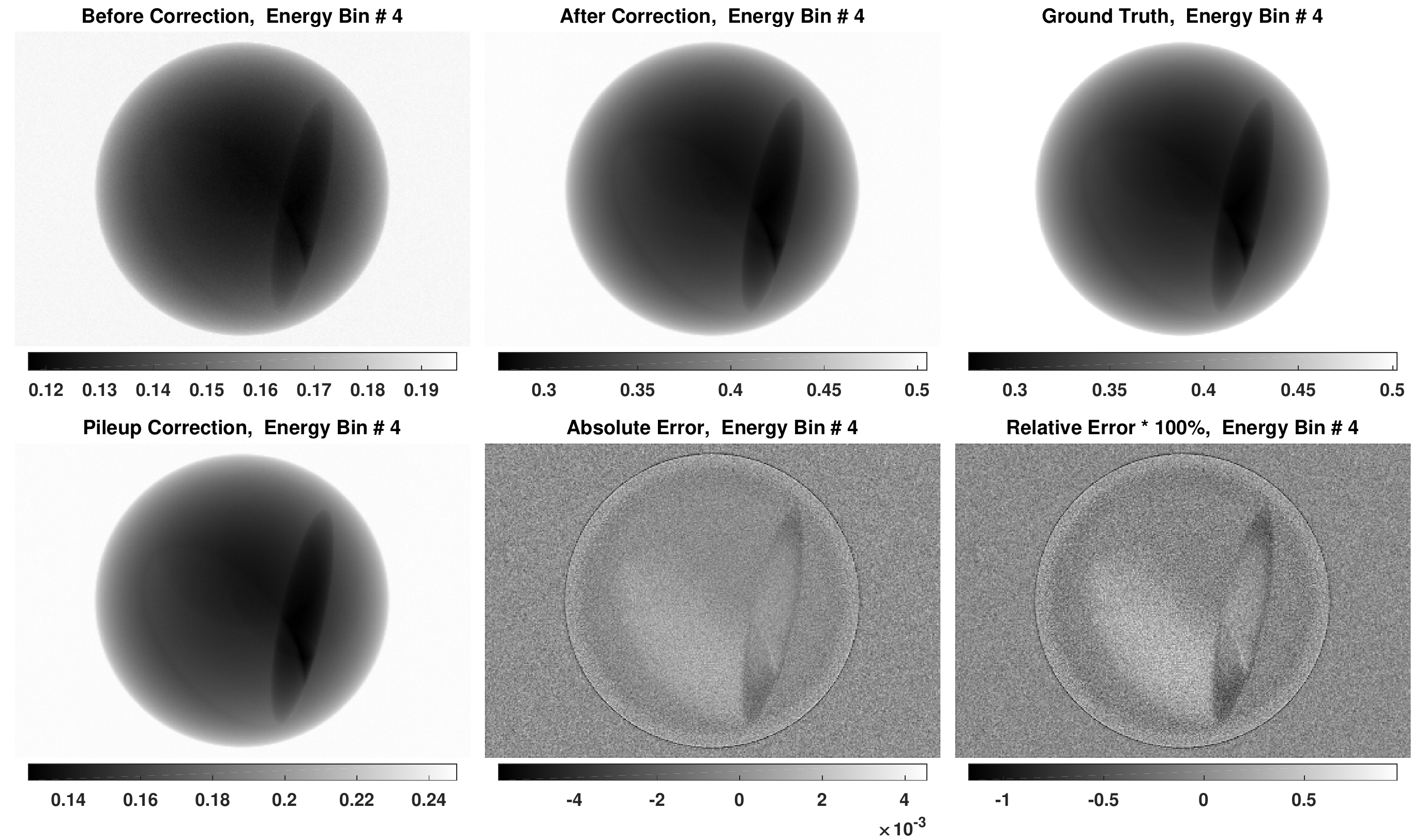}
    \caption{Projection image at $180^\circ$ angle view before and after correction in energy bin 4 ($50-59keV$). Counts are normalized with 400,000.}
    \label{fig:Proj4}
\end{figure}

\begin{figure}
    \centering
    \includegraphics[width=0.8\textwidth]{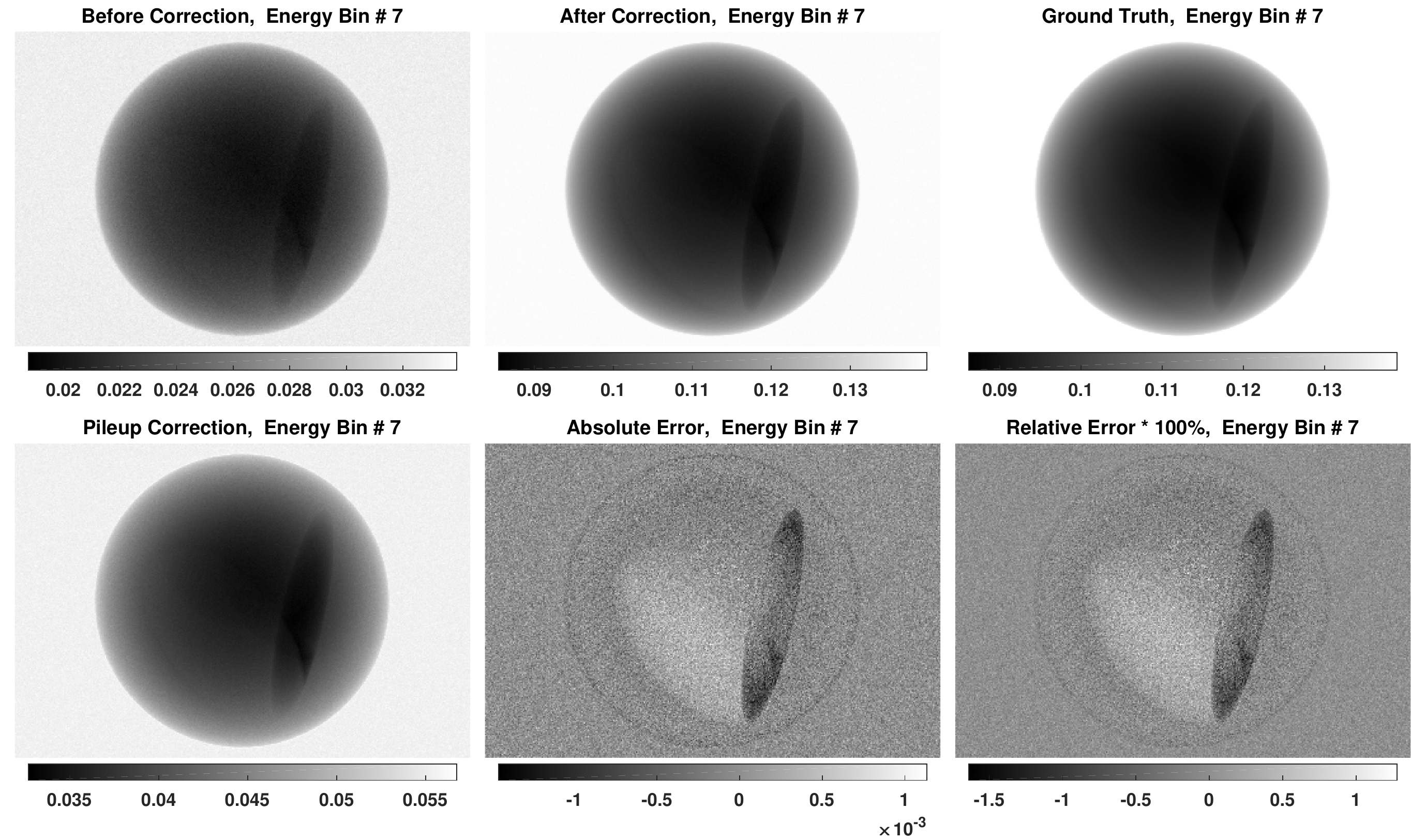}
    \caption{Projection image at $180^\circ$ angle view before and after correction in energy bin 7 ($70-79keV$). Counts are normalized with 400,000.}
    \label{fig:Proj7}
\end{figure}

To demonstrate the data fidelity improvement with our method, the trained network was applied on the test data, and two energy channel images (bin 4 and bin 7) of one representative projection view with rich structures are shown in Figs. \ref{fig:Proj4} and \ref{fig:Proj7}. It is hard to visually discern any difference between the corrected results and the ground truths. Quantitatively, the absolute errors are on the order of $10^{-3}$ in both figures, and the corresponding relative errors are negligible and smaller than $\pm1.5\%$ demonstrating the effectiveness of the network. Comparing the projections before and after correction, two major noticeable differences are the scales and the noise level besides a slight contrast improvement of structures after correction. Specifically, the scale range is around $[0.12,0.195]$ before correction compared to $[0.275, 0.5]$ after correction which is also the the range of the ground truth as shown in Fig. \ref{fig:Proj4}, demonstrating our method is indeed doing the correct transformation, and similar scale boost is seen in Fig. \ref{fig:Proj7}. In addition, the denoising effect is evidenced by the cleaner look of corrected projections compared to the raw measurement, especially in Fig. \ref{fig:Proj7} which suffers serve Poisson noise due to less counts in the energy bin, and demonstrates apparent noise reduction. 

\begin{figure}
   \begin{minipage}[b]{.49\linewidth}
     \centering
     \includegraphics[width=0.8\textwidth]{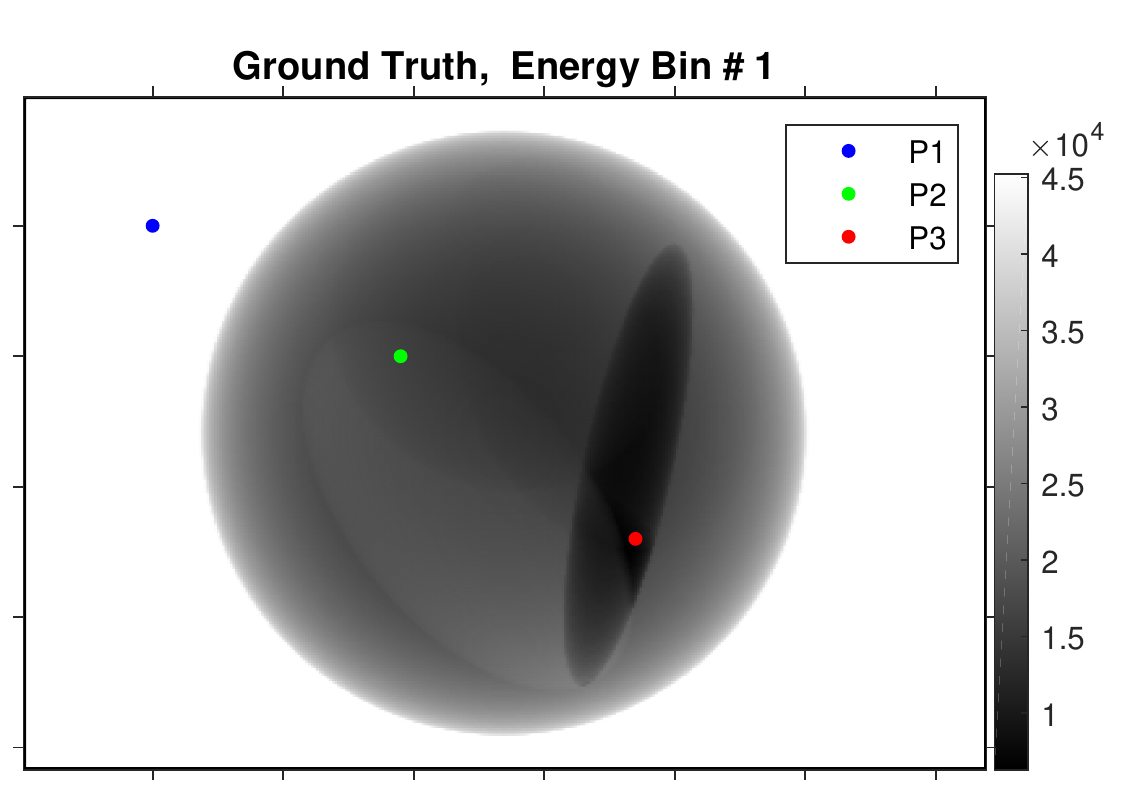}
     \subcaption{Point positions illustration}\label{fig:Points_illustration}
   \end{minipage}
   \hfill
   \begin{minipage}[b]{.49\linewidth}
     \centering
     \includegraphics[width=0.9\textwidth]{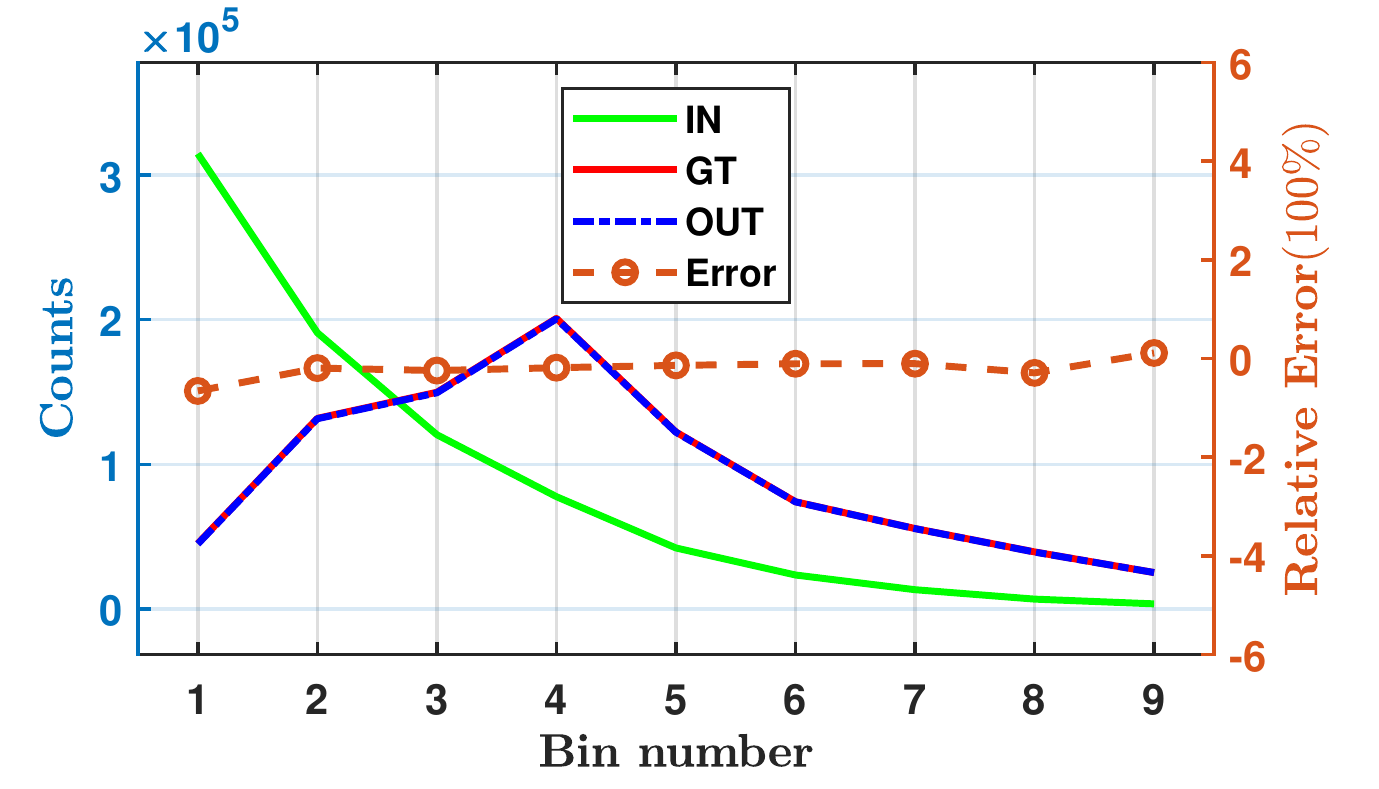}
     \subcaption{Energy profiles of P1}\label{fig:Correction_profile1}
   \end{minipage}
   \begin{minipage}[b]{.49\linewidth}
     \centering
     \includegraphics[width=0.9\textwidth]{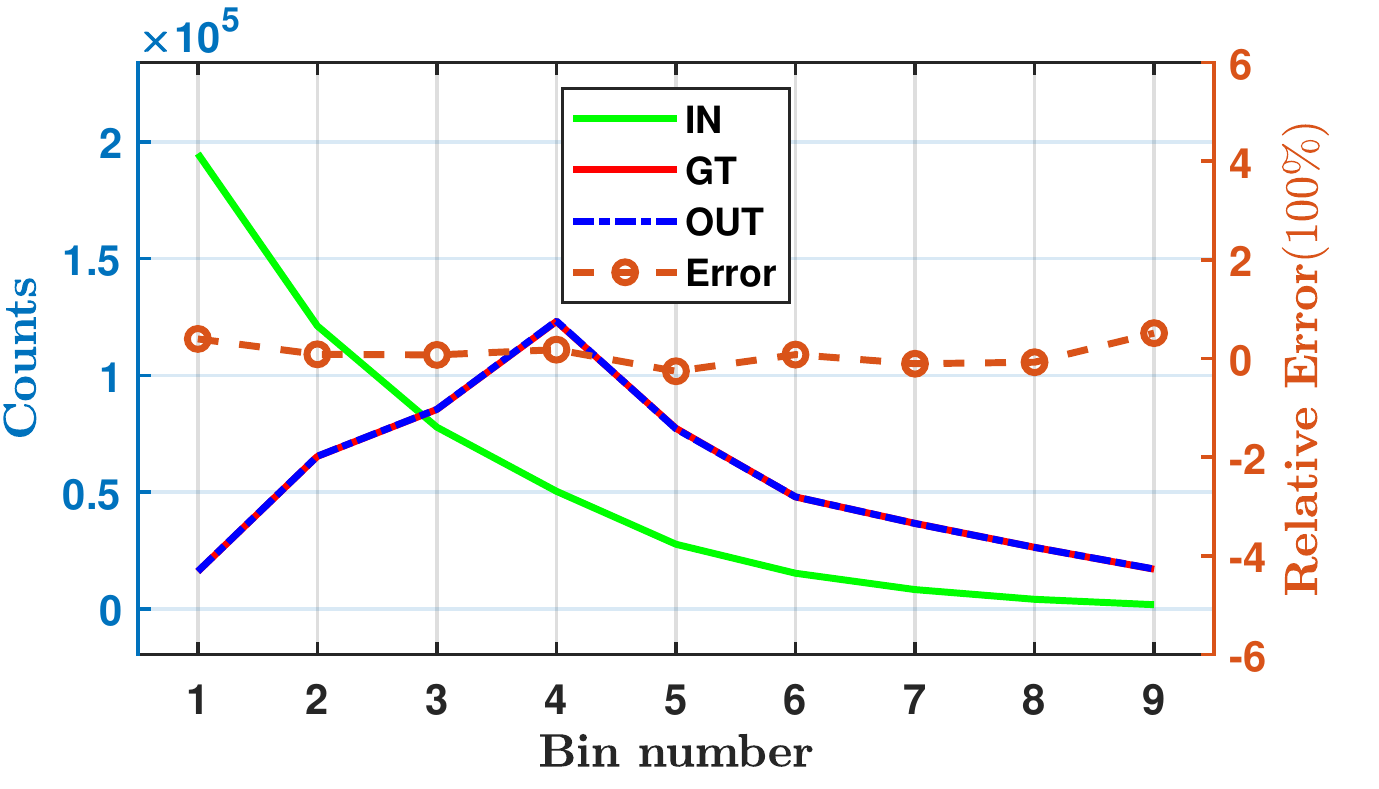}
     \subcaption{Energy profiles of P2}\label{fig:Correction_profile2}
   \end{minipage}
   \hfill
   \begin{minipage}[b]{.49\linewidth}
     \centering
     \includegraphics[width=0.9\textwidth]{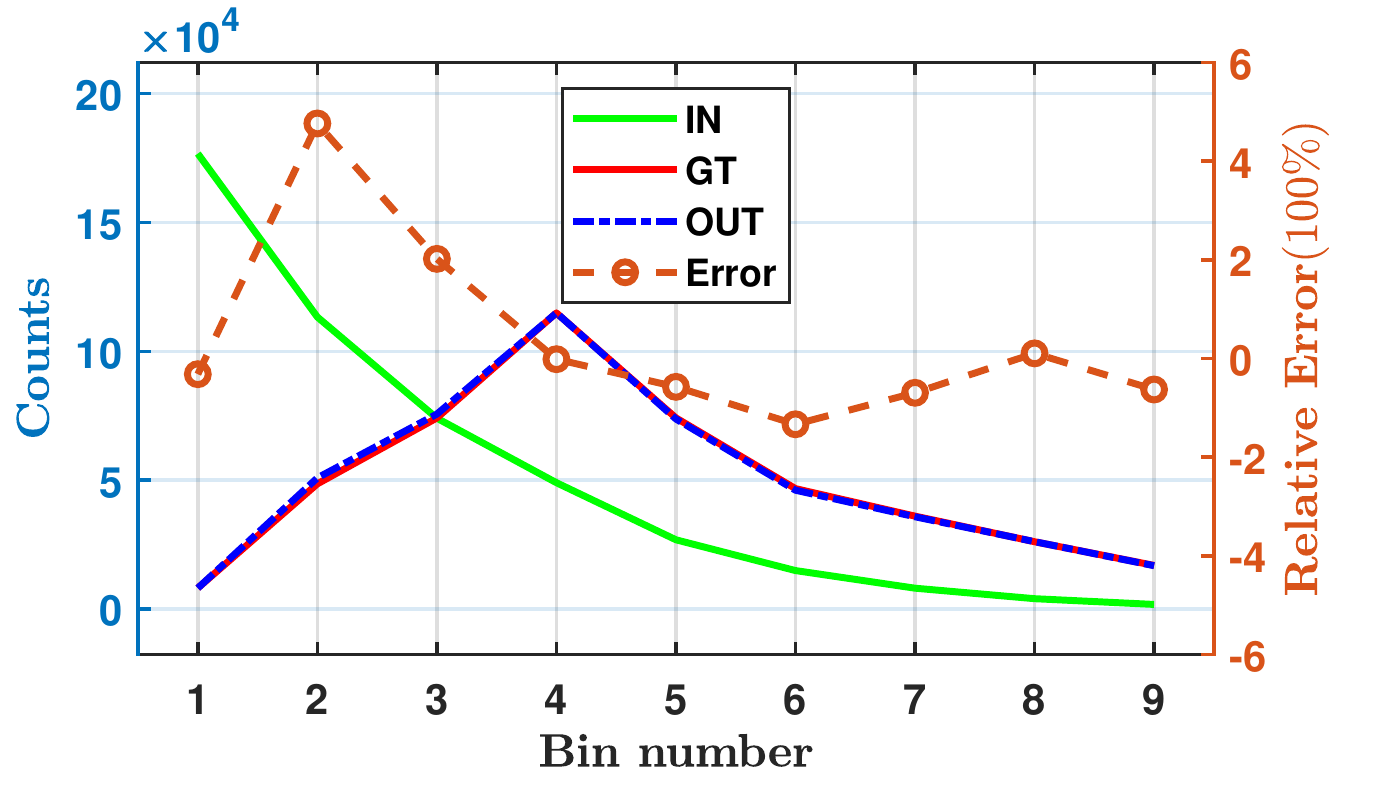}
     \subcaption{Energy profiles of P3}\label{fig:Correction_profile3}
   \end{minipage}
   \caption{Energy profiles of three representative points before and after correction. Corresponding point positions of (b),(c) and (d) are illustrate in (a). IN: input profile before correction; GT: ground truth profile; OUT: output profile of the correction network; Error: relative error calculated by (OUT-GT)/GT.}
   \label{fig:count_correction}
\end{figure}

Energy profiles in all energy bins were studied with three representative points experiencing different levels of attenuation, i.e., air, tissue and bones, as shown in Fig~\ref{fig:count_correction}.
The point positions are illustrated in Fig. \ref{fig:Points_illustration}, which displays the background is the ground truth projection at $180^\circ$ angle view in energy bin 1 ($20-29keV$). From Figs. \ref{fig:Correction_profile1}, \ref{fig:Correction_profile2} and \ref{fig:Correction_profile3}, it is clear that the significantly distorted input spectra are transformed to the correct shapes by the network, and the profiles after correction are almost overlapped perfectly with the ground truths.
From the relative error curves, we can find that the correction errors are within $\pm 1\%$ for median or less attenuation cases in all energy bins, and for the heavy attenuation case shown in Fig. \ref{fig:Correction_profile3}, the error is within $\pm 5\%$ for all bins and less than $\pm 2\%$ for energy higher than $40keV$. 

\begin{figure}
    \centering
    \includegraphics[width=0.4\textwidth]{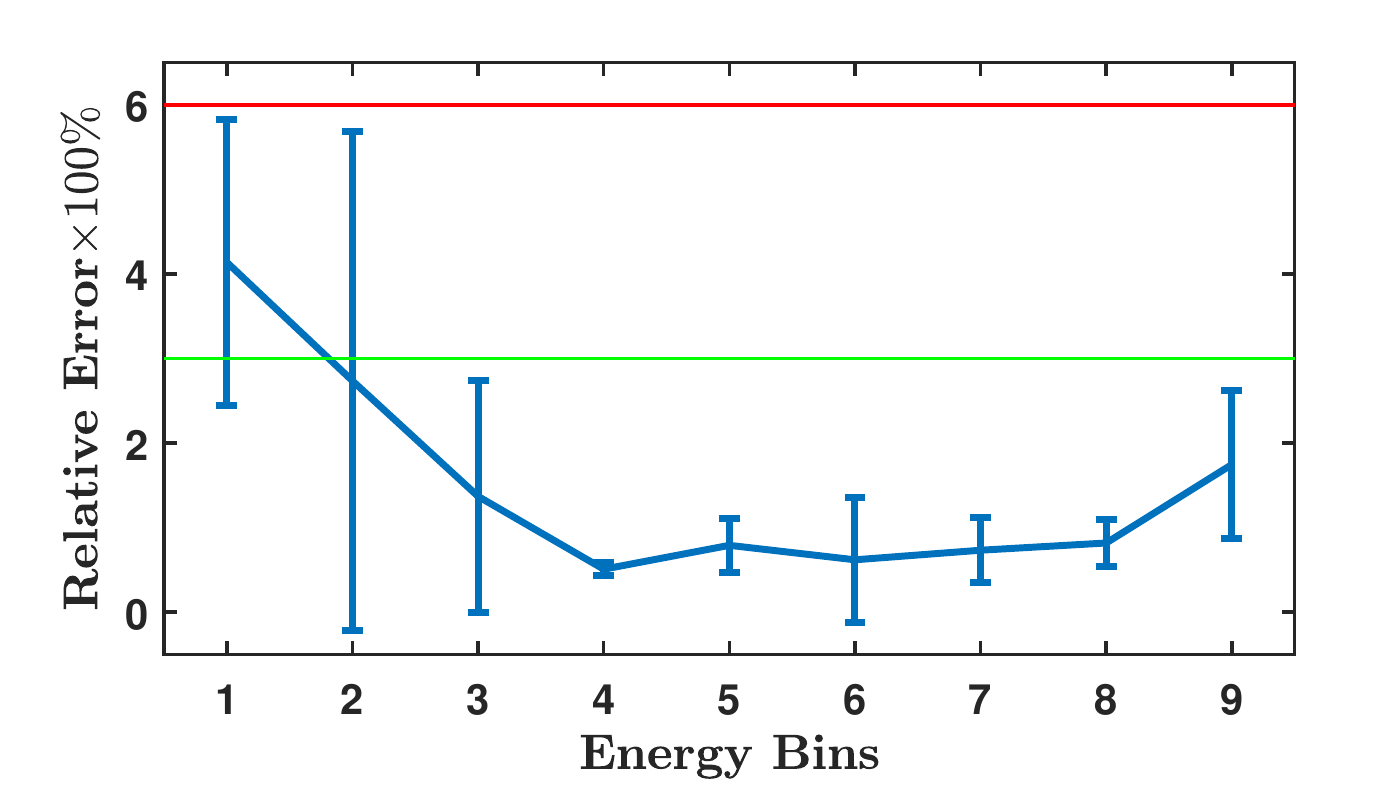}
    \caption{Mean $99\%$ percentile relative error curve with 3 folds of std. errorbar.}
    \label{fig:CrtErrAssessment}
\end{figure}

To further investigate the correction accuracy, 5 testing datasets each with 180 projection views were used to assess the relative errors range.  First, we found the maximum relative error in each projection view for each energy bin, and obtained $180\times5$ datapoints for each energy bin; Then, the maximum value, mean value and standard deviation (std.) were calculated out of the absolute values of these datapoints; Finally, we followed the same procedures for the percentiles of $99\%$, $95\%$ and $90\%$ respectively. The results are shown in Table \ref{table:Accuracy}. The first row in the table implies that the maximum relative error for the whole test data is controlled below $55\%$ for all energy bins after correction, which is impressive compared to the original value over $1000\%$ indicated in Fig.~\ref{fig:Correction_profile3}. Comparing the columns, the bins 3 to 8 demonstrate much smaller mean values of the maximum, even smaller than $5\%$, compared to the bins 1, 2 and 9. This is reasonable due to the fact that bins 1 and 2 are in the more complicated situations and are most likely to receive counts of fluorescence and charge-sharing from neighbor pixels, while bin 9 has less counts which suffers more from Poisson noise. Besides the extreme cases including some abnormal pixels, studying the majority of the pixels is also meaningful which dominantly decide the reconstruction quality. The mean values of the $99\%$ percentile relative errors from every view are all smaller than $5\%$ in all bins. More visually, the mean value curve with error-bars of 3 folds of standard deviation is plotted in Fig.~\ref{fig:CrtErrAssessment}, and it can be found that all errors with error-bars are below the red $6\%$-line and those from bins 3 to 9 are even below the green $3\%$-line. Suppose the errors follow normal distributions, the results suggest that $99\%$ pixels have been corrected to the ground truths within $6\%$ relative error for all energy bins ($20keV$ to $110keV$) at a $99.7\%$ confidence, and within $3\%$ for bins 3 to 9 ($40keV$ to $110keV$).

\begin{table}[htbp]
\begin{center}
\caption{\label{table:Accuracy} Correction accuracy on five testing datasets}
\begin{tabular}{lcccccccccc}
\toprule 
\multicolumn{2}{c}{{Relative Error ($100\%$)}} &  Bin 1 & Bin 2 & Bin 3 & Bin 4 & Bin 5 & Bin 6 & Bin 7 & Bin 8 & Bin 9 \\
\midrule
\multirow{3}{4em}{{Maximum}}
& Max & 54.997 & 32.321 &  7.337 &  2.055 &  3.936 & 12.725 &  3.479 &  2.688 & 29.405\\
& Mean & 12.335 &  7.066 &  3.151 &  1.177 &  1.892 &  1.859 &  1.825 &  1.912 &  4.930\\
& Std. & 5.043 &  3.698 &  1.082 &  0.135 &  0.345 &  1.506 &  0.389 &  0.286 &  2.813\\
\hline
\multirow{3}{4em}{{$99\%$ pct.}}
& Max & 5.967 &  4.696 &  2.403 &  0.576 &  1.106 &  1.355 &  1.119 &  1.029 &  2.720\\
& Mean & 4.137 &  2.735 &  1.369 &  0.508 &  0.790 &  0.619 &  0.733 &  0.818 &  1.745\\
& Std. & 0.565 &  0.985 &  0.457 &  0.026 &  0.106 &  0.246 &  0.129 &  0.094 &  0.293\\
\hline 
$95\%$ pct. & Max & 2.936 &  1.428 &  0.759 &  0.401 &  0.564 &  0.383 &  0.502 &  0.598 &  1.124\\
\hline
$90\%$ pct. & Max & 2.298 &  1.055 &  0.561 &  0.323 &  0.430 &  0.268 &  0.378 &  0.453 &  0.838\\
\bottomrule
\end{tabular}
\end{center}
\end{table}

\begin{figure}
    \centering
    \includegraphics[width=0.8\textwidth]{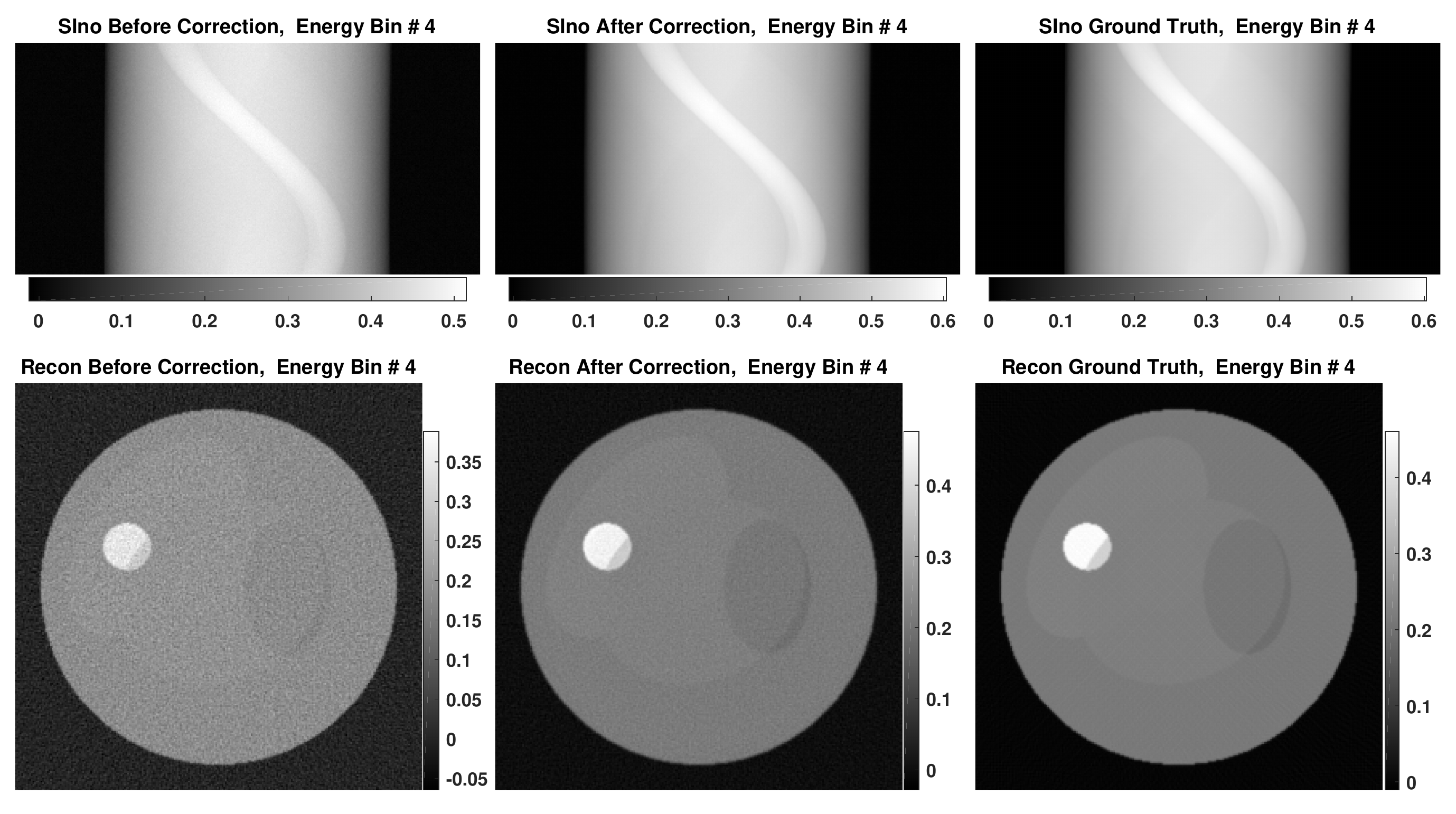}
    \caption{Sinograms and reconstructions of data in bin 4 ($40-49keV$) before and after correction comparison. Units in $cm^{-1}$}
    \label{fig:ReconSino4}
\end{figure}

\begin{figure}
    \centering
    \includegraphics[width=0.8\textwidth]{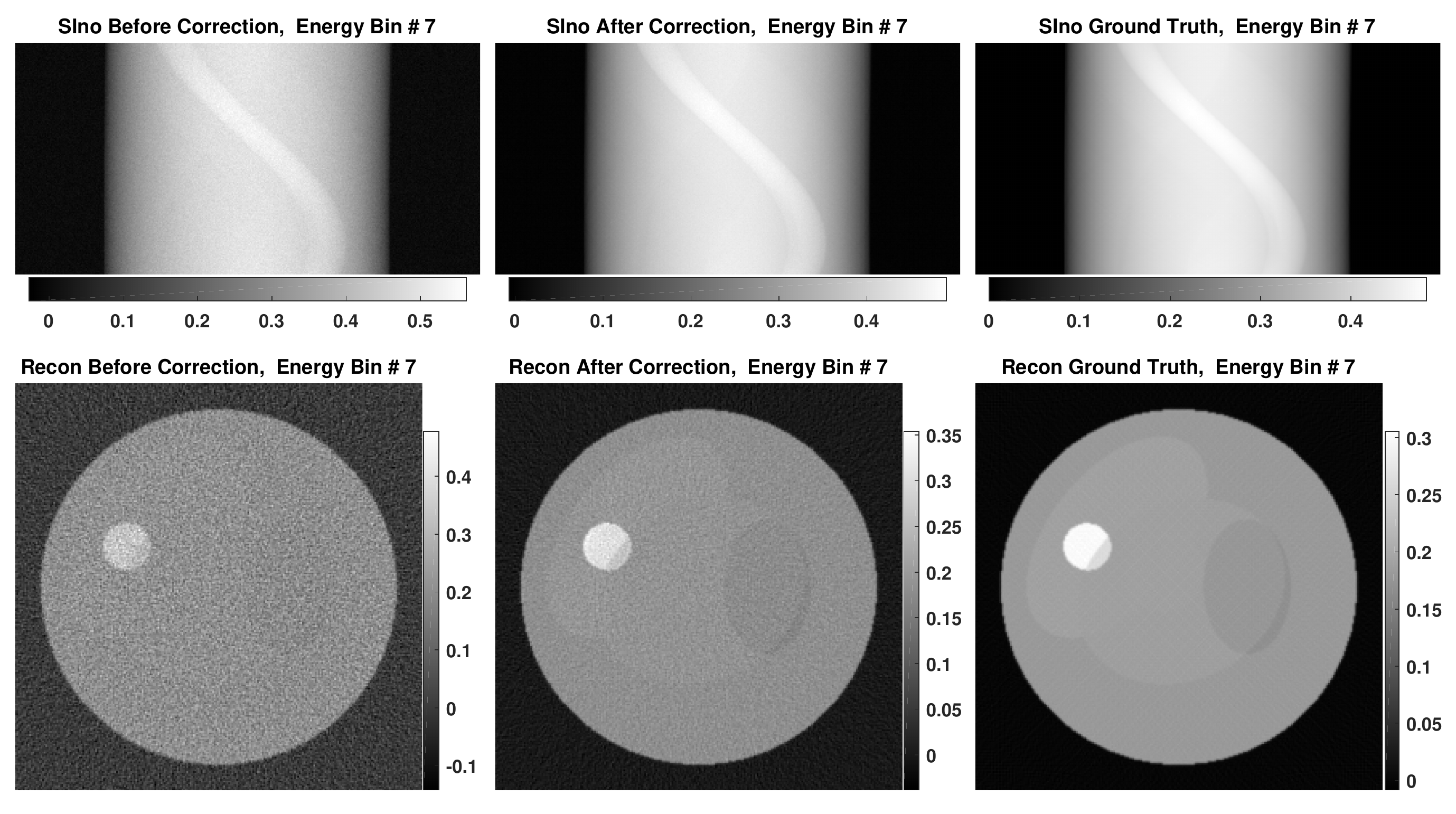}
    \caption{Sinograms and reconstructions of data in bin 7 ($70-79keV$) before and after correction comparison. Units in $cm^{-1}$}
    \label{fig:ReconSino7}
\end{figure}

\begin{figure}
   \begin{minipage}[b]{.3\linewidth}
     \centering
     \includegraphics[width=1\textwidth]{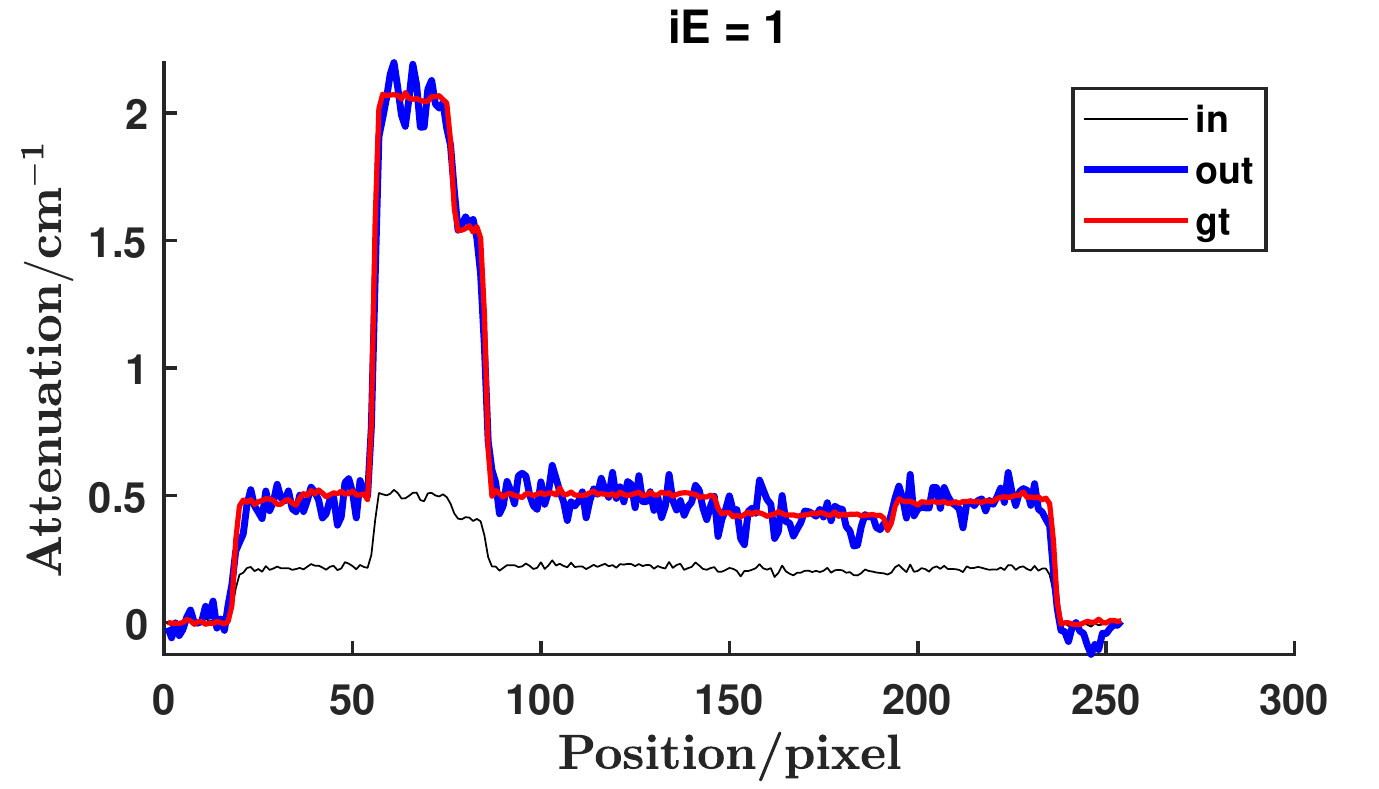}
     \subcaption{Attenuation profiles in bin 1}\label{fig:AttProfile1}
   \end{minipage}
   \hfill
   \begin{minipage}[b]{.3\linewidth}
     \centering
     \includegraphics[width=1\textwidth]{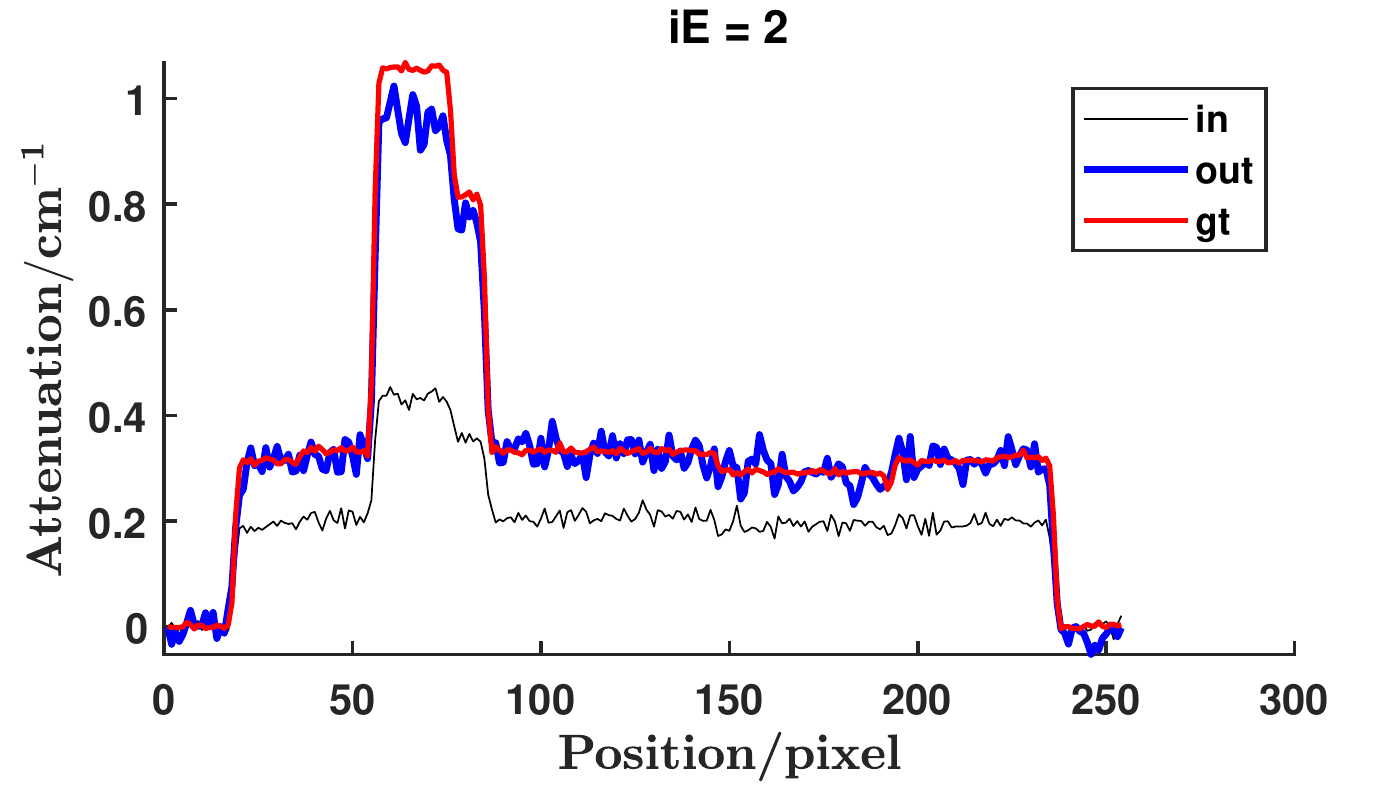}
     \subcaption{Attenuation profiles in bin 2}\label{fig:AttProfile2}
   \end{minipage}
   \hfill
   \begin{minipage}[b]{.3\linewidth}
     \centering
     \includegraphics[width=1\textwidth]{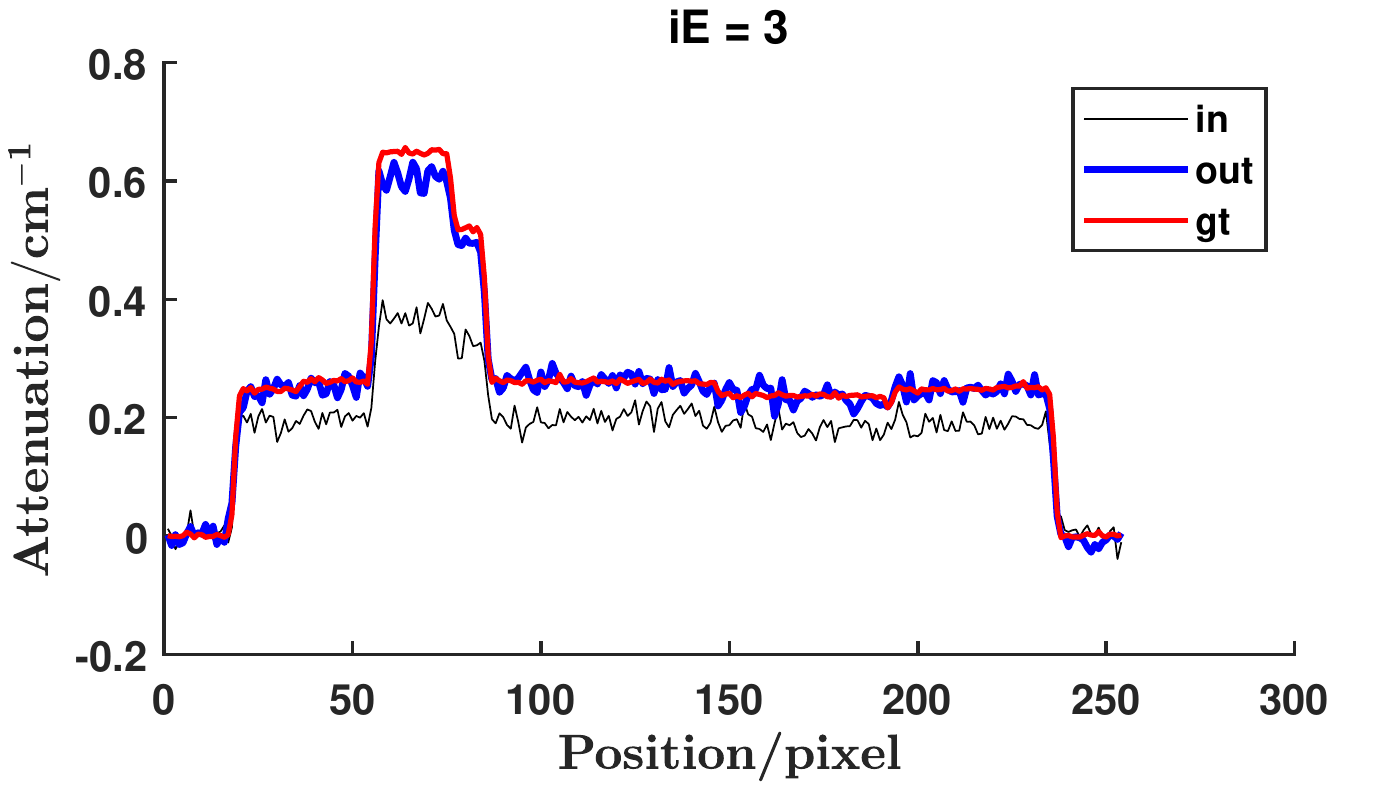}
     \subcaption{Attenuation profiles in bin 3}\label{fig:AttProfile3}
   \end{minipage}

   \begin{minipage}[b]{.3\linewidth}
     \centering
     \includegraphics[width=1\textwidth]{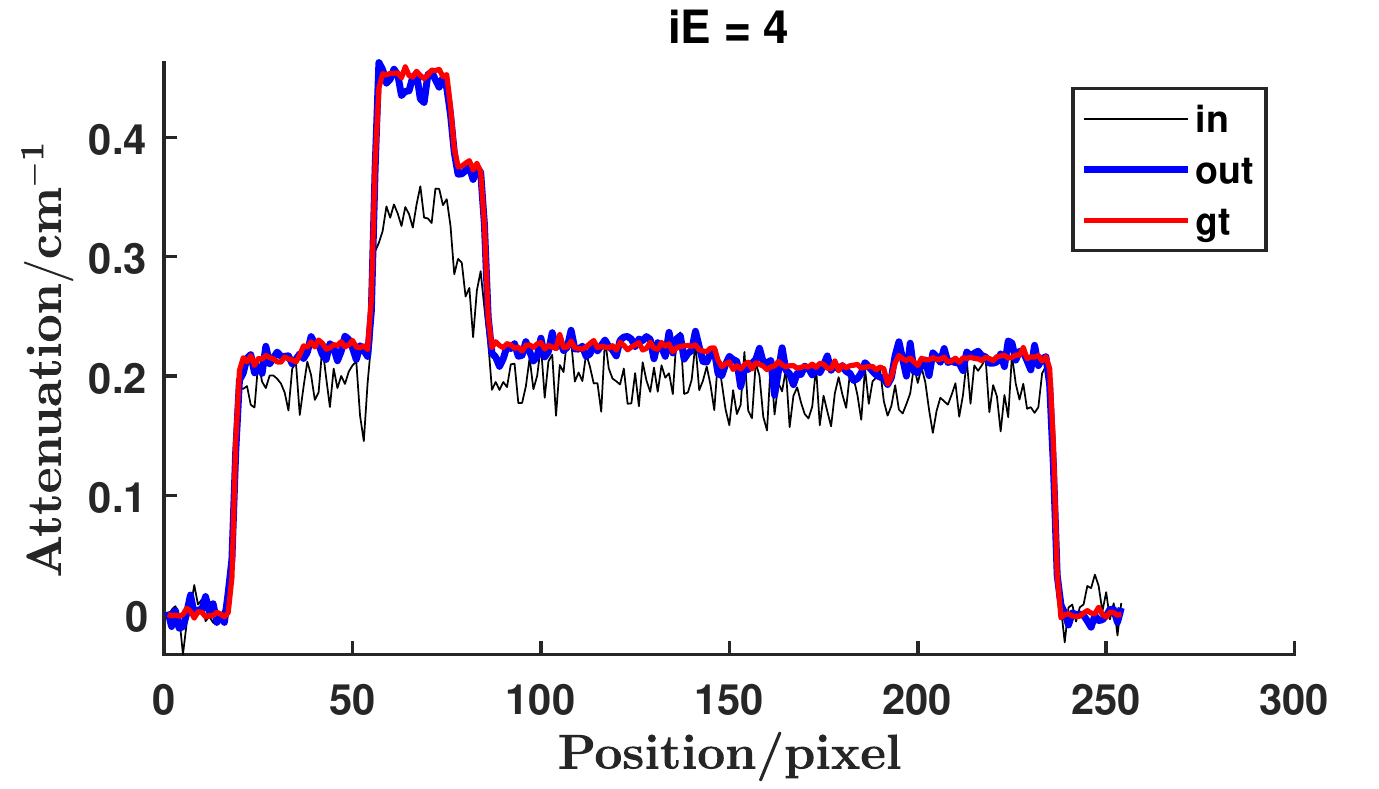}
     \subcaption{Attenuation profiles in bin 4}\label{fig:AttProfile4}
   \end{minipage}
   \hfill
   \begin{minipage}[b]{.3\linewidth}
     \centering
     \includegraphics[width=1\textwidth]{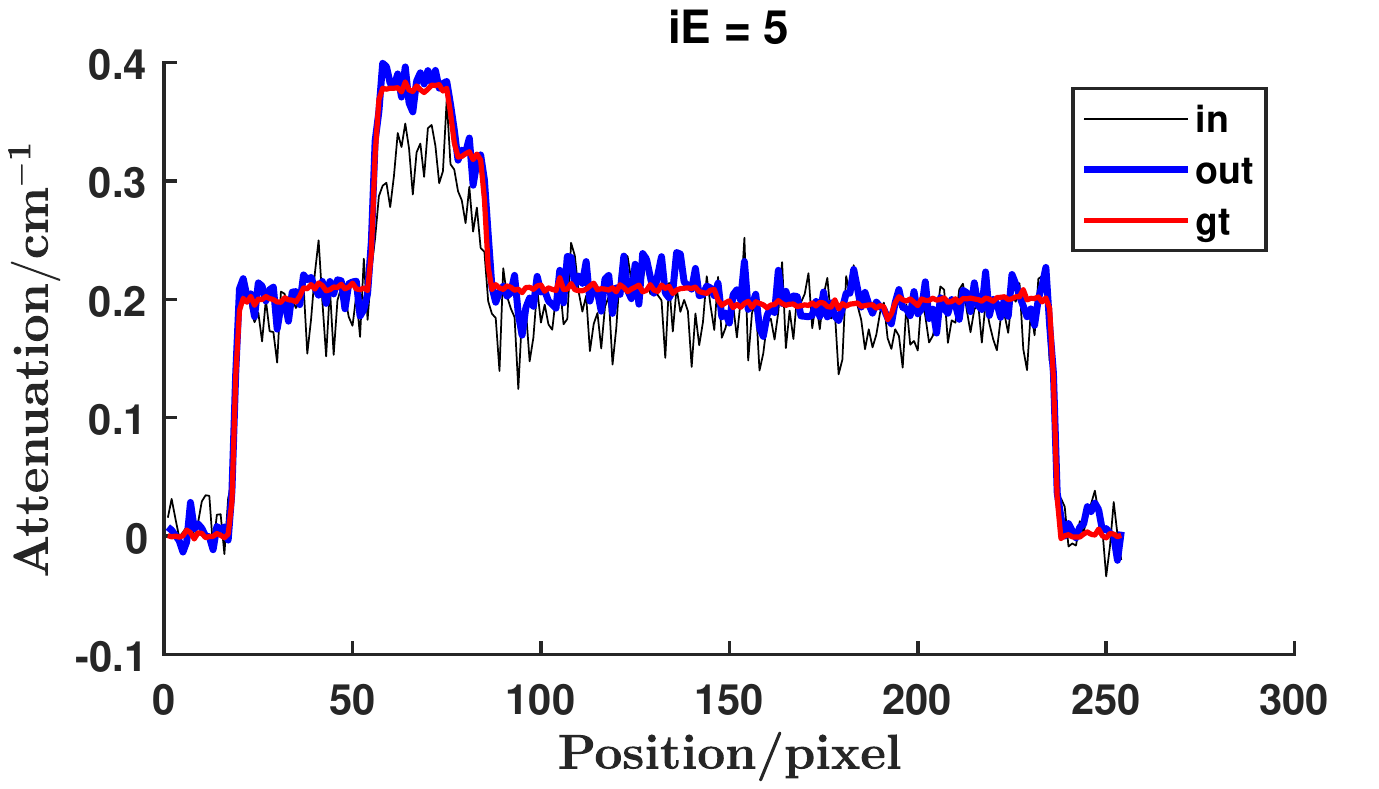}
     \subcaption{Attenuation profiles in bin 5}\label{fig:AttProfile5}
   \end{minipage}
   \hfill
   \begin{minipage}[b]{.3\linewidth}
     \centering
     \includegraphics[width=1\textwidth]{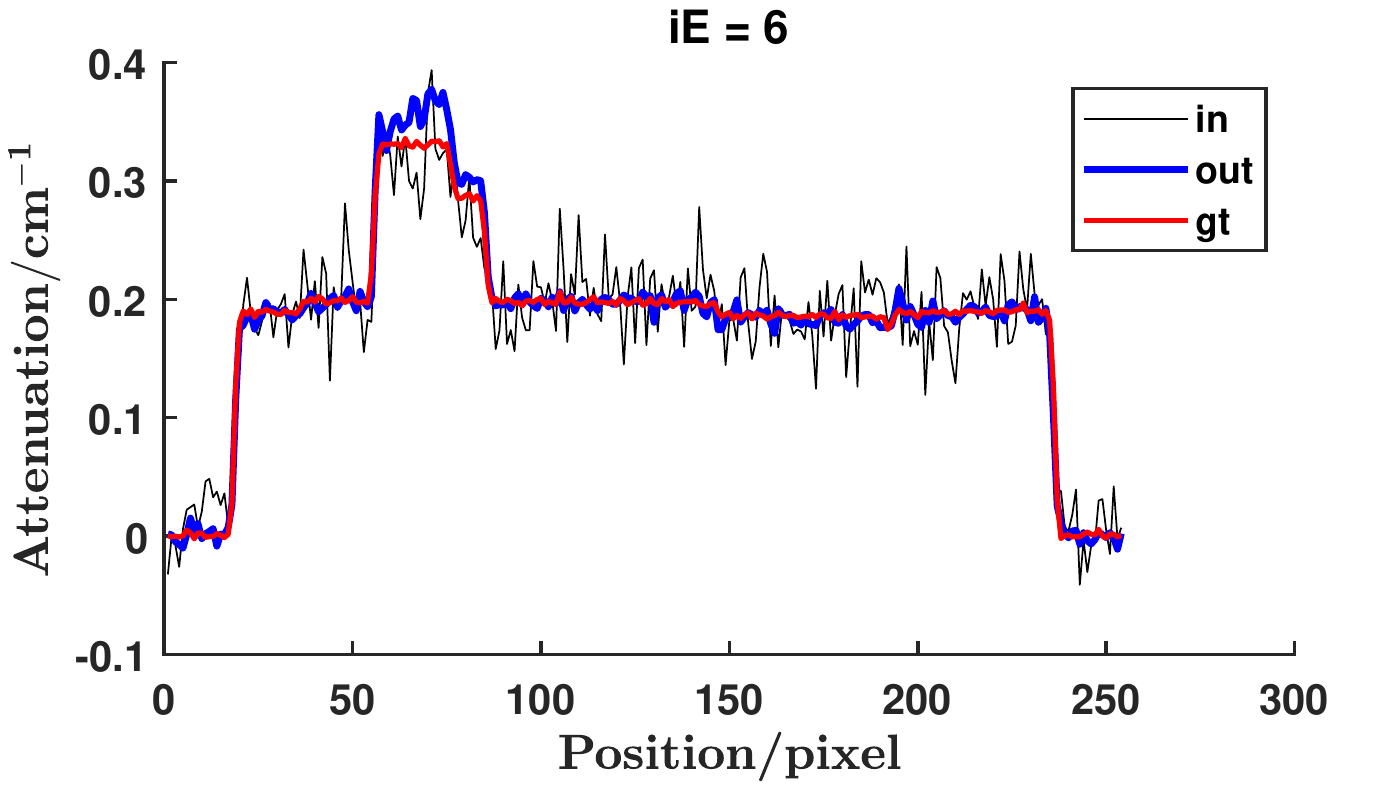}
     \subcaption{Attenuation profiles in bin 6}\label{fig:AttProfile6}
   \end{minipage}

   \begin{minipage}[b]{.3\linewidth}
     \centering
     \includegraphics[width=1\textwidth]{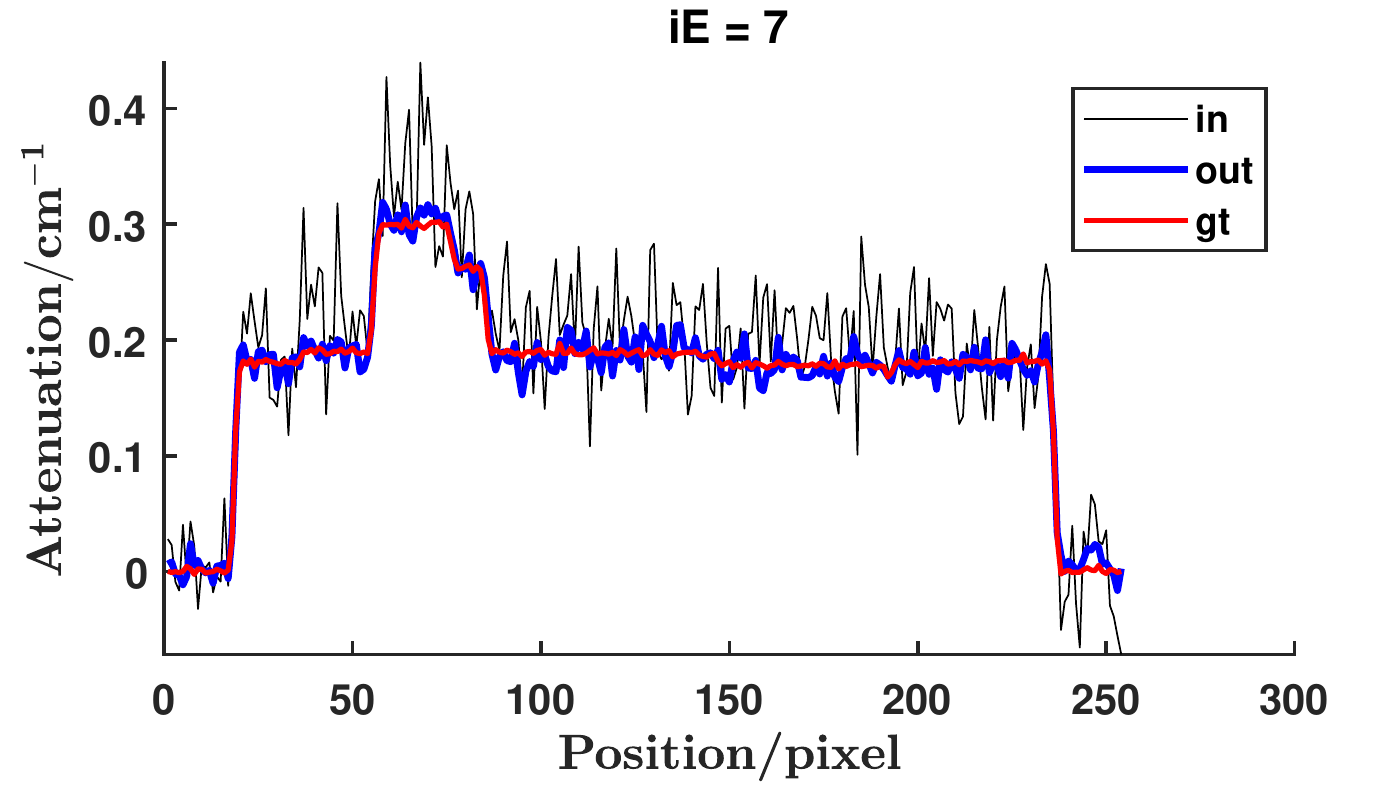}
     \subcaption{Attenuation profiles in bin 7}\label{fig:AttProfile7}
   \end{minipage}
   \hfill
   \begin{minipage}[b]{.3\linewidth}
     \centering
     \includegraphics[width=1\textwidth]{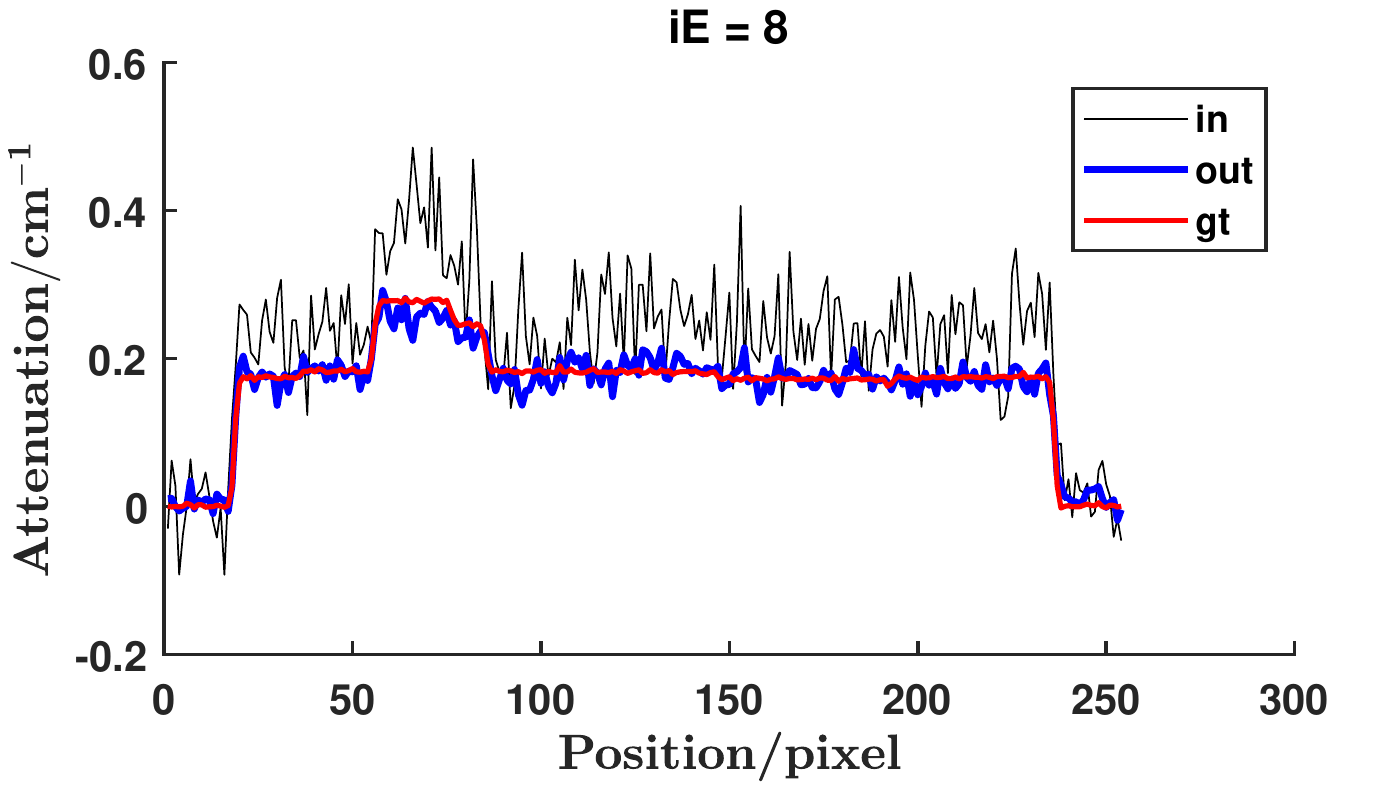}
     \subcaption{Attenuation profiles in bin 8}\label{fig:AttProfile8}
   \end{minipage}
   \hfill
   \begin{minipage}[b]{.3\linewidth}
     \centering
     \includegraphics[width=1\textwidth]{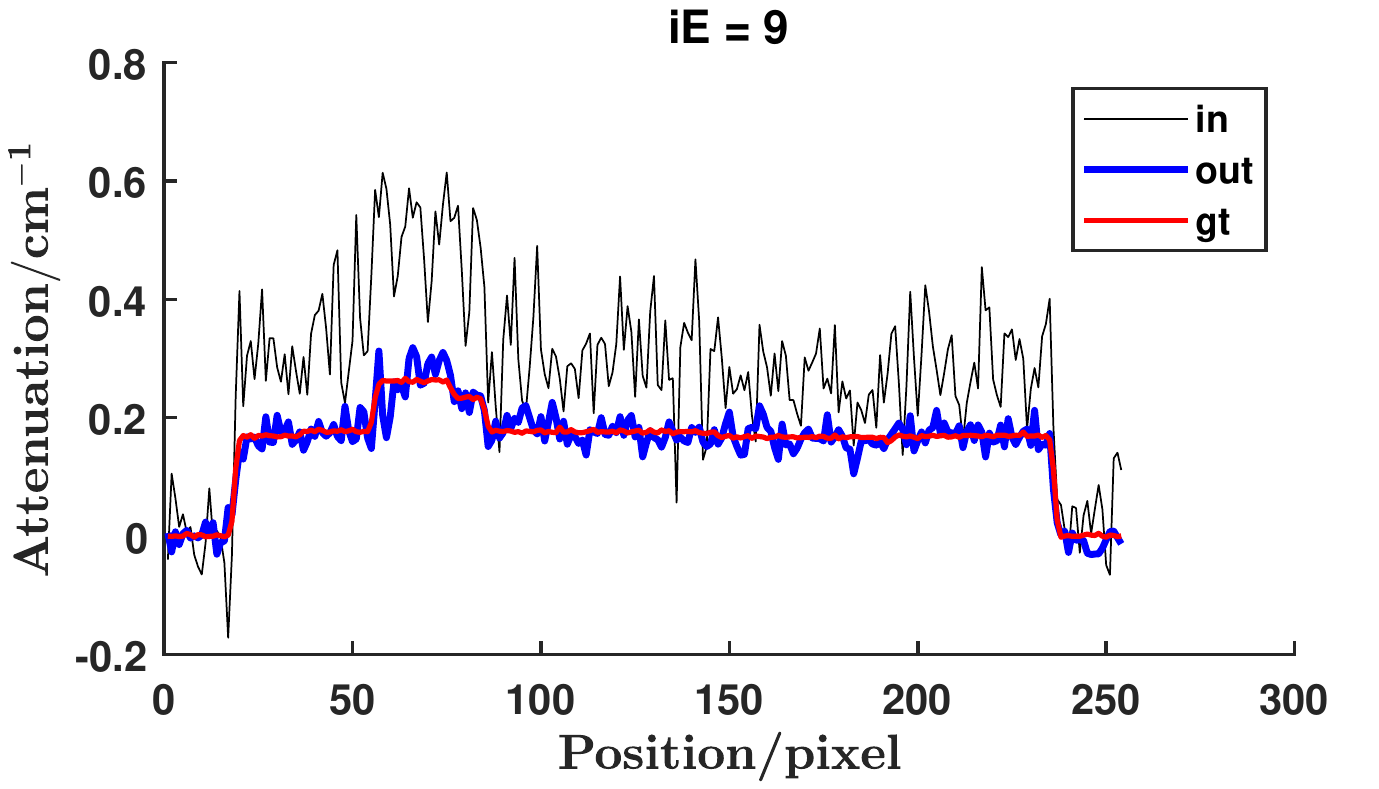}
     \subcaption{Attenuation profiles in bin 9}\label{fig:AttProfile9}
   \end{minipage}
   \caption{Attenuation profiles ($105_{th}$ row) in the reconstruction before and after correction, where IN denotes input profile before correction, GT stands for ground truth profile, and OUT represents output profile of the correction network.}
   \label{fig:AttProfiles_all}
\end{figure}

To cast light on the correction effects on the reconstruction, sinograms with air correction were calculated from the data before and after spectral correction, and those sinograms were then reconstructed using filtered back-projection (FBP) method. The reconstructions and corresponding sinograms of a representative slice with rich structures (the $95_{th}$ row of the projections in Figs.~\ref{fig:Proj4} and \ref{fig:Proj7}) in bins 4 and 7 are shown in Figs. \ref{fig:ReconSino4} and \ref{fig:ReconSino7}, respectively. Similar scale changes described in projection images are observed here in terms of LAC before and after correction, which implies the effectiveness of correction improving the reconstruction accuracy. Moreover, obvious noise reduction and contrast improvement are demonstrated in both figures, and especially in Fig. \ref{fig:ReconSino7}, the structures immersed in noise before correction are successfully recovered and noise in air region is strongly suppressed. Comparing the correction results with the ground truths, they are almost the same in terms of structures despite residual noise in correction results. To assess the accuracy of LAC values after correction, line profiles of the $105_{th}$ row of the reconstruction in all energy bins are plotted in Fig. \ref{fig:AttProfiles_all}. Overall, the profiles after correction follow the ground truths very well, except for slight regional mismatches in bins 2, 3 and 6, in contrast to the results before correction which deviate considerably from the ground truths especially at high or low energy bins and accompanied by increasingly growing noise from low to high energy bins. To be mentioned, the drop-down of LAC before correction in the low energy bins is due to the contamination of the charge-splitting counts which represent the LAC in higher energy bins, while the lift-up in the high energy bins is mainly contributed by the pulse-pileup counts which represent the LAC in lower energy bins and the pileup effect will also additionally raise the calculated LAC. In summary, the correction demonstrates huge improvement in both LAC value accuracy and noise level as well as image contrast. 


For quantitative assessment, structural similarity index (SSIM) \cite{wang2004image} and peak signal-to-noise ratio (PSNR) metrics have also been evaluated on reconstructions before and after correction with reference to the ground truths, and listed in Table \ref{table:QM}. Not surprisingly, the results after correction score significantly higher than those before correction in both metrics. Especially, the improvements are more prominent in low and high energy bins, e.g., the SSIM score and PSNR score are more than doubled after correction in bin 1 and bin 9 respectively. An interesting phenomenon that the SSIM score before correction first rises up and then falls down as energy increases results from the large deviation of the LAC values from the ground truths in both low and high energy bins demonstrated in Fig. \ref{fig:AttProfiles_all}. Similar trends are observed in the PSNR scores of uncorrected results, and they drop faster in high energy bins due to the monotonous increase of noise. Bins 4 to 6 get good scores before correction because the spectral distortion does not change the structures while the LAC deviation and noise level are relatively small in the middle energy range as shown in Fig. \ref{fig:AttProfiles_all}.

\begin{table}[htbp]
\begin{center}
\caption{\label{table:QM} Quantitative metrics on the reconstruction slice shown in Figs. \ref{fig:ReconSino4} and \ref{fig:ReconSino7}}
\begin{tabular}{lccccccccc}
\toprule 
GT as ref. &  Bin 1 & Bin 2 & Bin 3 & Bin 4 & Bin 5 & Bin 6 & Bin 7 & Bin 8 & Bin 9 \\
\midrule
Uncorrected (SSIM) & 0.4669 & 0.7679 & 0.9114 & 0.9636 & 0.9691 & 0.9562 & 0.9172 & 0.8265 & 0.6633 \\
Corrected (SSIM) & \bf0.9835 & \bf0.9896 & \bf0.9943 & \bf0.9977 & \bf0.9931 & \bf0.9976 & \bf0.9936 & \bf0.9890 & \bf0.9698 \\
\hline
Uncorrected (PSNR) &  19.99 &  21.58 &  23.47 &  25.43 &  25.34 &  22.96 &  19.01 &  14.19 &   9.09\\
Corrected (PSNR) & \bf 33.63 & \bf 34.20 & \bf 35.10 & \bf 37.41 & \bf 31.76 & \bf 35.75 & \bf 31.04 & \bf 28.36 & \bf 23.73\\
Improvement ($100\%$) & 68.2 & 58.5 & 49.6 & 47.1 & 25.3 & 55.7 & 63.3 & 99.9 & 161.1 \\
\bottomrule
\end{tabular}
\end{center}
\end{table}

\section{DISCUSSIONS AND CONCLUSION}
We have proposed an end-to-end deep-learning-based approach for PCD data correction for high spectral fidelity, and also validated the method effectiveness with realistically synthesized PCD data. In the paper, a new PCD data simulation model was described, which incorporates two major phenomena, charge splitting (charge sharing and fluorescence) and pulse pileup, that contribute to the spectral response degradation. The charge splitting part is based on Ken's toolkit, and the  pulse pileup model is adapted from Roessl's analytical modeling method to incorporate the spatial cross-talk. New network structure dedicated for the correction task of PCD data has been proposed, which includes dePUnet and deCSnet targeted on pulse pileup correction and charge splitting correction respectively. 

The whole workflow of PCD data generation was described in the paper from realistic simulation of ideal spectral projection to PCD detection simulation. Those pairs of synthetic distorted measurements and corresponding ground truths were fed to the network for training. Then, experiments with additionally simulated testing data were conducted to evaluate the fidelity improvements after correction with the trained network in both projection and reconstruction domains. The proposed network demonstrates great denoising and spectral correction abilities from the visual test on two representative projections. The quantitative test on large test dataset shows that the maximum pixel-wise relative error has been controlled below $55\%$ over the whole dataset compared to the over-$1000\%$ before correction, and $99\%$ pixels have been corrected to the ground truth with errors smaller than $6\%$ for all energy bins at a $99.7\%$ confidence, and with errors less than $3\%$ for bins 3 to 9 corresponding to the energy range from $40keV$ to $110keV$. From the perspective of reconstruction, the correction demonstrates obvious noise-suppression effect and huge LAC fidelity improvement as expected in the experiment. In terms of quantitative metrics, the correction improves the SSIM scores by $0.024\sim 0.517$ and the PSNR score by $25.3\% \sim 161.1\%$. Those excellent testing results strongly evidence the promising performance of deep learning application in PCD data correction. 

The limitations of the work mainly come from the PCD data simulation model, such as: (1) No Compton scattering included; (2) Cross-talks only limited to neighboring $3\times3$ pixels while the size should be larger when small pixels are used; (3) Charge trapping no included; (4) Nonuniformity of pixels not included. Those minor effects are not modeled but indeed exist during the real PCD detection. Thus, the procedures to apply the proposed approach in practice should involve real measurement data, and could be (1) directly collecting the distorted PCD data and calculating corresponding ideal spectral projections as the ground truth first, and then training the network using those data pairs; Or (2) estimating the electronic noise parameter $\sigma_e$ and the effective electron clouds radius $r_0$ of the real PCD first, then generating the simulation data to train the network to avoid tedious data collection process, and finally using a small amount of real PCD measurement for transfer learning. It is worth noting that Touch presented PCD correction work with similar goals through combination use of a shallow fully-connection network for spectral distortion correction and a joint bilateral filtration denoising method\cite{touch2016neural}. In contrast, our approach is an end-to-end method which achieves denoising and distortion correction simultaneously through the correction network, and our network adopts the state-of-the-art deep learning techniques and has powerful representation ability while their network only contains two hidden layer and each with only five neurons. We also trained their network with the same data for comparison but the MSE loss during training only went down to 0.12 after convergence while ours reached down to 0.0000011. The huge difference suggests that their network might be overwhelmed by the various spectrum shapes induced by the variations in attenuation length and material bases, and failed this task. Thus, we did not perform further tests on their network.

In conclusion, we have proposed a deep-learning-based PCD data correction approach for spectral fidelity which includes a new PCD data simulator with both pulse pileup and charge splitting effects modeled, and a specially designed network dedicated for PCD data correction. The testing results with synthetic data suggest the proposed approach achieved obvious noise suppression and accurate spectral correction in both projection and reconstruction domains. The ideal spectrum shape are faithfully recovered from the significantly distorted measurement within $\pm6\%$ relative errors in all bins at a $99.7\%$ confidence. The SSIM and PSNR scores between the reconstruction and the ground truth are greatly improved. In the future, we plan to collect real PCD spectral projection data, implement the workflow outlined above, and evaluate the performance.

\bibliographystyle{spiebib} %

\end{document}